\documentclass[journal,final,letter,onecolumn,12pt]{IEEEtran}

\usepackage{graphicx}
\usepackage{amsmath,bbm}
\usepackage{epsfig}%
\usepackage{subfigure}
\def\bz{{\bf z}}
\def\CC{{\cal C}}

\def\CK{{\cal K}}
\def\bu{{\bf u}}
\def\bv{{\bf v}}
\def\bw{{\bf w}}
\def\ie{{\em i.e., }}
\def\eg{{\em e.g., }}
\def\A{{\cal A}}
\def\TP{\tilde{P}}

\newtheorem{theorem}{Theorem}
\newtheorem{corollary}{Corollary}
\newtheorem{lemma}{Lemma}

\newtheorem{remark}{Remark}

\begin{document}
\title{On Space-Time Capacity Limits in Mobile and Delay Tolerant Networks
\thanks{Part of this work will be presented in ``On Space-Time Capacity Limits in Mobile and Delay Tolerant Networks'', P. Jacquet, B. Mans and G. Rodolakis, IEEE Infocom, 2010.}
}

\author{
Philippe Jacquet, Bernard Mans and Georgios Rodolakis
\thanks{%
P. Jacquet is with INRIA, 78153 Le Chesnay, France. E-mail: philippe.jacquet@inria.fr}
\thanks{%
B. Mans and G. Rodolakis are with Macquarie University, 2109 NSW, Australia.
E-mails: bernard.mans@mq.edu.au, \mbox{georgios.rodolakis@mq.edu.au}}
}

\maketitle
\begin{abstract}
We investigate the fundamental capacity limits
of space-time journeys
of information in mobile and Delay Tolerant Networks (DTNs),
where information is either
transmitted or carried by mobile nodes, using store-carry-forward routing.
We define the capacity of a journey (\ie a path in space and time, from a source to a destination) as
the maximum amount of data that can be transferred from the source to the destination in the given journey.
Combining a stochastic model (conveying all possible journeys) and an analysis of the durations of the nodes' encounters,
we study the properties of journeys that maximize the space-time information propagation capacity, in bit-meters per second.
More specifically, we provide theoretical lower and upper bounds on the information propagation speed, as a function of the journey capacity.
In the particular case of random way-point-like models (\ie when nodes move for a distance of the order of the network domain size before changing direction), we show that, for relatively large journey capacities, the information propagation speed is of the same order as the mobile node speed.
This implies that, surprisingly, in sparse but large-scale mobile DTNs, the space-time information propagation capacity in bit-meters per second remains proportional to the mobile node speed and to the size of the transported data bundles, when the bundles are relatively large.
We also verify that all our analytical bounds are accurate in several simulation scenarios.
\end{abstract}

\baselineskip 19 pt

\section{Introduction}
\label{Sect:intro}

The problem of determining fundamental limits on the
performance of mobile and ad hoc networks continues to attract
the interest of researchers. Several important results have been achieved with the seminal papers by Gupta and Kumar~\cite{GK00} (which provided the first capacity bounds in static wireless networks) and by Grossglauser and Tse~\cite{GT01} (which showed that the mobility can increase the capacity of an ad~hoc network).
Various mobility
models
have been studied in the literature, and the delay-capacity
relationships under those models have been characterized (\eg \cite{EMPS04,NM05,TG04}).
However, the nature of these trade-offs is strongly
influenced by the choice of the mobility model~\cite{SMS06}.

Moreover, there has been an increased interest in
mobile ad hoc networks where end-to-end multi-hop
paths may not exist and communication routes may only be available
through time and mobility; depending on the context, these networks
are now commonly referred as Intermittently Connected Networks
(ICNs) or Delay Tolerant Networks (DTNs).
Although limited,
the understanding of the fundamental properties of such networks  is steadily increasing.
There is a significant number of results focusing on characterizing the packet propagation delay~\cite{PMCC07,GNK05,ZNKT07},
assuming that packet transmissions are instantaneous,  and more recently, the information propagation speed~\cite{JMR09,KY08,KY08infocom}.
The authors of~\cite{PMCC07} took a graph-theoretical approach in order to upper bound the time it takes for disconnected mobile networks to become connected through the mobility of the nodes.
The papers~\cite{GNK05,ZNKT07} analyze the delay of common routing schemes, such as epidemic routing, under the assumption that the inter-meeting time between pairs of nodes follows an exponential distribution.
However, this assumption 
is not generally verified, depending on the relationship between the size of the network domain and the relevant time-scale of the network scenario under consideration~\cite{CY07}, and this can result in either an over-estimation or an under-estimation of the actual system performance~\cite{CY09}.
Departing from the exponential inter-meeting time hypothesis, in~\cite{KY08,KY08infocom}, Kong and Yeh studied the information dissemination latency in large wireless and mobile networks, in constrained i.i.d. mobility and Brownian motion models. They showed that, when the network is not percolated, 
the latency scales linearly with the Euclidean distance between the sender and the receiver.
The first analytical estimates of the constant upper bounds on the speed at which information can propagate
in DTNs, again without considering the quantity of information that can be transmitted,
were obtained in~\cite{JMR09}.

In contrast, in this paper, we investigate the space-time capacity of such networks,
\ie the maximum amount of information that can be transferred from a source to a destination over time.
As the network is almost surely disconnected, we refer to journeys rather than paths, where a journey is an alternation of data transmissions and carriages using store-carry-forward routing.
Informally, our objective is to determine how fast a given amount of data $y$ can reach its destination.
Formally, we use a probabilistic model of space-time journeys of packets of information in DTNs (in Section~\ref{Sect:model}), and define the journey capacity as well as the information propagation speed (in Section~\ref{Sect:thickness}), to provide the following main contributions:
\begin{itemize}
\item we characterize the duration of node meetings,
by bounding the probability function of the durations of the nodes' encounters,
in Section~\ref{Sect:meetings};
\item we prove the first non trivial
lower bounds on the information propagation speed (Theorem~\ref{theo:lower}), for a bounded journey capacity, in random waypoint-like mobility, in Section~\ref{Sect:lower};
\item
we prove general upper bounds on the information propagation speed (Theorem~\ref{theo:upper} and Corollaries~1 and~2), as a function of the journey capacity,
and we investigate the properties of journeys that maximize the space-time network capacity in bit-meters per second,
in Section~\ref{Sect:upper};
\item we compare and verify the analytical bounds with simulation measurements in Section~\ref{Sect:simulations}.
\end{itemize}
We provide concluding remarks in Section~\ref{Sect:conclusion}.

\section{Network and Mobility Model}
\label{Sect:model}

We consider a network of $n$ nodes in a square area of size
$\A=L\times L$ and radio range $R$.
As we want to focus on DTNs that are almost surely disconnected,
we will analyze the case where $R$ is fixed, while $n,\A \to \infty$, such that the node density $\nu=\frac{n}{\A}$ is bounded by some constant.

Formally, we adopt the random geometric graph model~\cite{P03}: two nodes at distance
smaller than a maximum radio range $R$ can exchange information.
Moreover, we consider that the rate at which nodes can transmit data when they are within range is fixed, and equals $G$ units of data per second.

Initially, the nodes are distributed uniformly at random.
Every node follows an i.i.d. random trajectory, reflected on the
borders of the square (like billiard balls).
The nodes change direction at Poisson rate~$\tau$ and keep a
constant speed $v$ between direction changes. The motion direction angles are
uniformly distributed in $[0,2\pi)$
and are mutually independent among all nodes.
When $\tau = 0$, we have a \emph{pure billiard model} (nodes only change direction at the border). 
When $\tau > 0$, we have a \emph{random walk model}; when $\tau\to\infty$ we are on the Brownian limit.
When $\tau = O(\frac{1}{L}) \to 0$ we
are on a \emph{random way-point-like model}, since nodes travel a distance of order $L$ before changing direction
or hitting the border.

\section{Space-Time Journey Analysis}
\label{Sect:thickness}

We study {\em journeys with a
given
capacity},
\ie journeys that guarantee that at least an amount of data
can be transferred to the destination.
Our aim is to find the shortest journey (in time)
with journey capacity at least $y$,
that
connects any source to any destination in the network domain, in order to derive
the overall
information propagation speed.

We base our analysis
on a probabilistic model of journeys of packets of information that encapsulates
all possible shortest journeys originating at the source, as used in~\cite{
JMR09}.
Let $\CC$ be a simple journey (\ie a journey not returning to the same node twice). Let $Z(\CC)$ be the terminal point.
Let $T(\CC)$ be the time at which the journey terminates. Let $p(\CC)$
be the probability of the journey $\CC$.

Let $\zeta$ be an inverse space vector, \ie with components expressed in inverse distance units. Let $\theta$ be a scalar in inverse time units.
We denote by
$w(\zeta,\theta)$ the journey Laplace transform, defined for a domain definition for $(\zeta,\theta)$: 
$$
\begin{array}{rcl}
w(\zeta,\theta)&=&E(\exp(-\zeta\cdot Z(\CC)-\theta T(\CC)))\\
&=&\sum_{\CC}p(\CC)\exp(-\zeta \cdot Z(\CC)-\theta T(\CC)).
\end{array}
$$

We call $p(\bz_0,\bz_1,t)$ the normalized density of journeys starting from $\bz_0$
at time~0, and arriving at~$\bz_1$ before time~$t$:
\begin{small}
$$
p(\bz_0,\bz_1,t)=\frac{1}{R^2}\sum_{\|\bz_1 -Z(\CC)\|<R,T(\CC)<t}p(\CC)~.
$$
\end{small}

\vskip - 0.42cm

Let us consider that a bundle of information of $y$ bits is generated at $t=0$ on a node at coordinate
$\bz_0=(x_0,y_0)$. Let us initially consider a destination node which stays motionless at coordinate $\bz_1=(x_1,y_1)$; in this case,
$p(\bz_0,\bz_1,t)$ denotes the probability that the destination receives one bit of information
before time~$t$.
Now,
let us consider a moving destination node,
that is located at coordinate $\bz_1=(x_1,y_1)$, at time $t$. We denote $\bz=\bz_1 - \bz_0$.
Let $q(\bz,t,y)$ denote the probability that there exists a journey of capacity at least $y$ reaching the destination before time~$t$.

The {\em information propagation speed} $s(y)$, considering a journey capacity $y$, is defined as the minimum ratio of distance over time above which the journey probability tends to $0$, \ie
\begin{itemize}
\item
if~$\frac{||\bz||}{t} > s(y)$, then
$\lim_{||\bz||,t \to\infty} q(\bz,t,y)=0$;
\item if~$\frac{||\bz||}{t} <s(y)$, then
$\lim_{||\bz||,t \to\infty} q(\bz,t,y)>0$.
\end{itemize}

We also define the {\em space-time information propagation capacity} $c(y)$ (from now on simply referred to as the {\em space-time capacity}), as the maximal transport capacity in bit-meters per second, that can be achieved by any journey of capacity~$y$.
Thus, in this model, the space-time capacity corresponds to the product $c(y)=s(y) y$.

Therefore, in order to determine the space-time capacity limits of mobile and delay tolerant networks, we will analyze the information propagation speed, as a function of the journey capacity;
in the following sections, we will compute lower and upper bounds.
In order to derive the bounds, we first study
the characteristics of node meetings.

\section{Node Meetings}
\label{Sect:meetings}

A meeting (or encounter) between two nodes occurs when their distance becomes smaller than or equal to $R$, \ie
when the nodes come into communication range.

\begin{lemma}
A node $A$, moving in direction $\psi_0$, meets new nodes moving in direction between $\psi_1$ and $\psi_1 + d\psi$ at
rate:
$f_{\psi_1~|~\psi_0}=\frac{2 v \nu R}{\pi}\sin (\frac{\psi_1-\psi_0}{2}) d\psi$, for $\psi_0,~\psi_1 \in (-\pi,\pi]$,
where $R$ is the radio range.
\label{lem:meeting_rate}
\end{lemma}

\begin{proof}
See appendix.
\end{proof}

We denote the meeting duration by the random variable $T$.

\begin{lemma}
The probability $P(T>t)$ that a meeting has duration at least $t$ satisfies:
$$
P(T>t) \leq \min(1,\frac{\pi^2 R}{8vt}).
$$
\label{lem:meetingbound}
\end{lemma}
\begin{proof}
The average number of neighbors of any node is $\pi \nu R^2$.
From Lemma~\ref{lem:meeting_rate}, the rate at which a node meets new neighbors is~$f=\frac{8 v \nu R}{\pi}$.
Therefore, from the Little formula, the average meeting time (\ie the time that a node remains a neighbor) equals $\frac{\pi \nu R^2}{f}=\frac{\pi^2 R}{8v}$.
The proof follows by applying Markov's inequality.
\end{proof}

In the pure billiard model (\ie when $\tau=0$), we can give the exact formulas on the meeting time distribution.
We note that our model where nodes bounce on the borders like billiard balls is equivalent to considering an infinite area made of mirror images of the original network domain square: a mobile node moves in the original square while its mirror images move in the mirror squares~\cite{JMR09}.

\begin{lemma}
We denote the meeting duration by the random variable $T$.
The probability density function $p_T(t)$ of $T$ is:
\begin{equation}
p_T(t)=\frac{1}{4}\log\left| \frac{\frac{v}{R}t+1}{\frac{v}{R}t-1}\right| \left( 1+\frac{R^2}{(vt)^2}\right)-\frac{R}{2vt},
\label{eq:densT}
\end{equation}
for $t\geq 0$, where $v$ is the node speed, $R$ is the radio range.

When $t\to\infty$, the cumulative probability $P(T>t)$ is:
$$
P(T>t) = \frac{R^2}{3(vt)^2}+ O\left( \frac{R^4}{(vt)^4} \right).
$$
\label{lem:distT}
\end{lemma}

\begin{proof}
See appendix.
\end{proof}

\section{Lower Bound}
\label{Sect:lower}
We prove a lower bound $s_L(y)$ on the information propagation speed, for journey capacity $y$, in the random way-point-like mobility model, \ie when nodes travel a distance of the order of the network domain length before changing direction.
Initially, we focus on the pure billiard mobility model, \ie we assume that nodes do not change direction unless they hit the border. Finally, we remark that the result can be generalized to node mobility with a small change of direction rate.

We will show that, for all destination nodes which, at time $t$, are at distance $r\sim s_L(y) t$ of the initial source location, there is a journey of duration $t$ and of capacity $y$ from the source to the destination,
with probability strictly larger than~$0$.
We consider large distances $r=\Theta(\sqrt{n})$, where $n$ is the number of nodes in the network; in this case, the square network domain has a side length $r=\Theta(\sqrt{n})$, as we are interested in the case where the node density is constant
(but strictly larger than~0),
as discussed in Section~\ref{Sect:model}.
We show that,
when the journey capacity is $y \leq \frac{K}{v}$, for a constant $K$, the lower bound is $s_L(y)=v$, where $v$ is the mobile node speed.

\begin{figure}[th]
\begin{center}
\includegraphics[width=4.1cm]{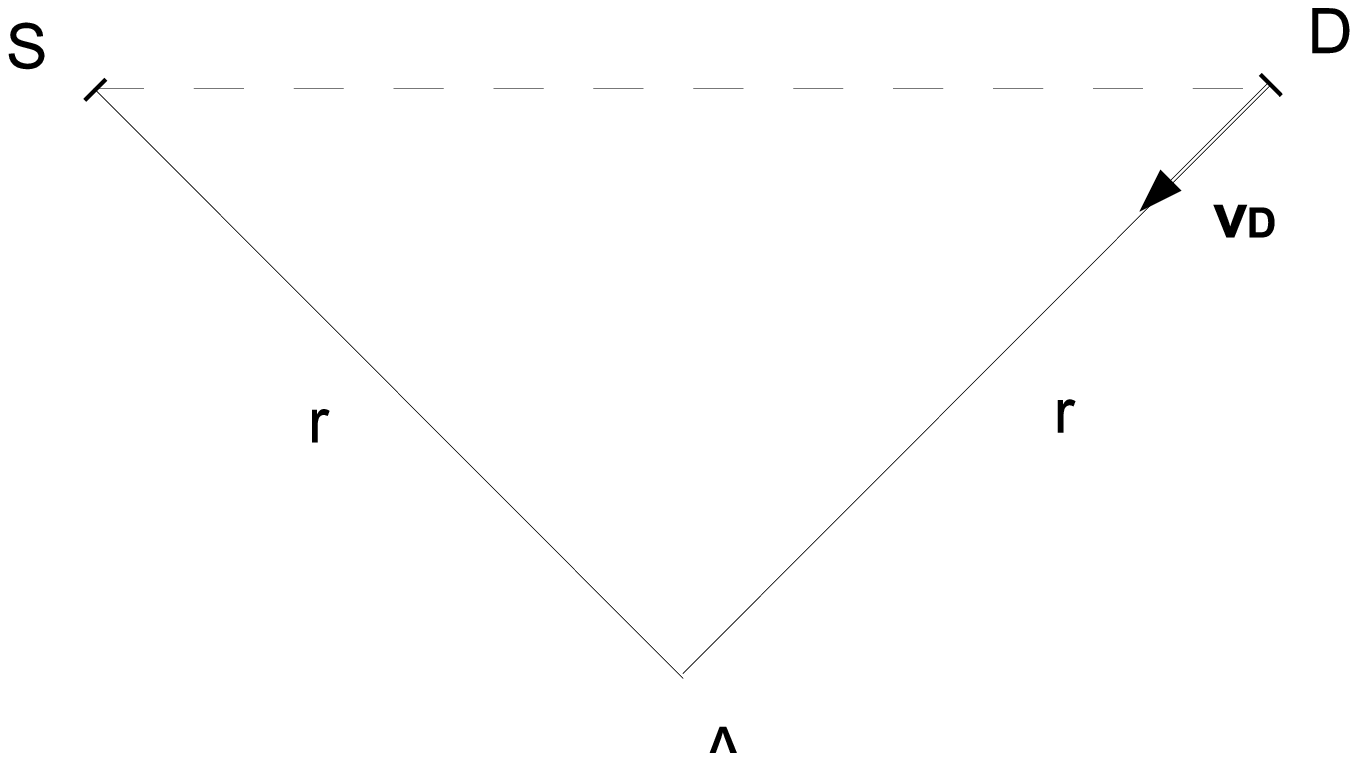}
\includegraphics[width=4.1cm]{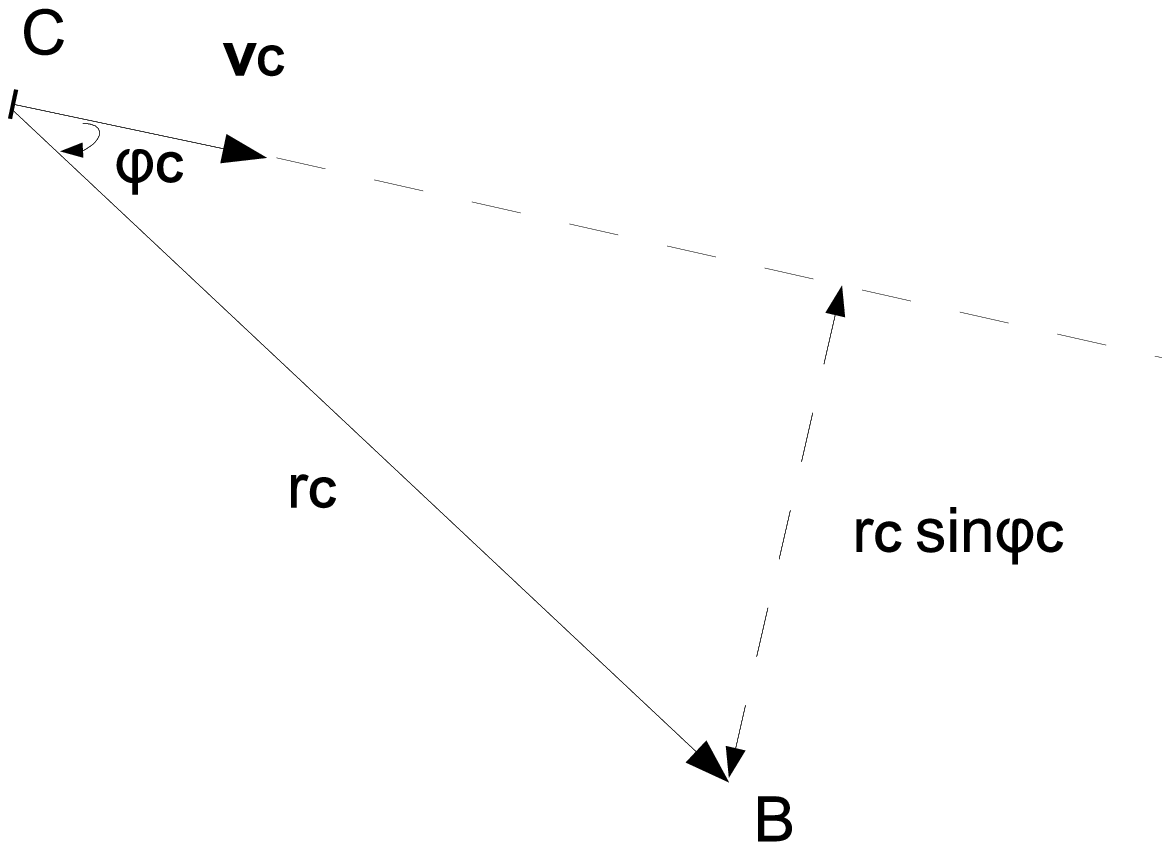}
\vskip -0.5cm
\caption{Definitions of rendez-vous point $A$ of the information generated at location $S$ with the destination $D$ (left), and of angle $\phi_C$ with respect to the speed of node $C$ and location $B$ (right).}
\label{meetingpoint}
\end{center}
\end{figure}

We consider a source node $S$ and a destination node $D$.
We denote by $\bv_S$ and $\bv_D$ the respective vector speeds of the source and the destination.
We assume that the source starts sending the information at time $0$.
We define the point $A$ as the third vertex of the isosceles triangle, formed with the two other vertices located at $S$ and $D$ (at time $0$) and with sides $SA$ and $DA$ of equal length $r$, while $DA$ is parallel to the destination speed $\bv_d$, as illustrated in Figure~\ref{meetingpoint}.
Point $A$ is therefore the \emph{rendez-vous point} of a node moving at constant speed $v$, in the direction of $SA$, and the destination node, while the nodes contact (at the same location) occurs at time $t_{A}=\frac{r}{v}$.
Similarly, if the (asymptotic) information propagation speed is equal to the node speed $v$, the information will reach the destination at location $A'=A \pm \Delta Z$, with $|\Delta Z|=o(r)$, at time $t_{A'}=t_A+o(\frac{r}{v})$.

We will describe a routing scheme that constructs a journey of duration $t_A = \frac{r}{v}+o(\frac{r}{v})$, which originates at $S$ and ends at any given point $A$, and guarantees that for any direction of the destination node speed, the journey capacity is at least~$y$. We assume w.l.o.g. that the radio range is $R=1$ and the communication rate is also $G=1$, to simplify the expressions (to generalize, it is sufficient to perform a simple scaling). We note that, in this case, ensuring a journey capacity at least equal to~$y$ is equivalent to ensuring a minimum meeting duration~$y$ for all transmissions in the journey.

\begin{figure}[t!]
\begin{center}
\includegraphics[width=6.5cm]{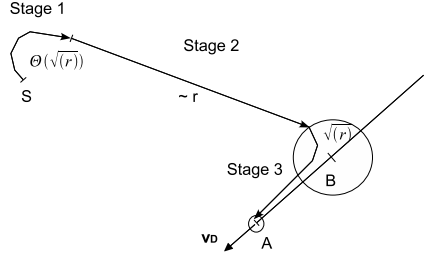}
\vskip -0.9cm
\caption{Overview of the routing scheme achieving the lower bound of information propagation towards the rendez-vous point $A$, in three stages.}
\label{lowerbound}
\end{center}
\end{figure}

The routing scheme proceeds in three stages, illustrated in Figure~\ref{lowerbound}. In all stages, the information is passed among nodes moving at relative direction of angle between $\frac{a}{2}$ and $a$, with a value of $a$ that we will precise in the following.

Initially, we consider a point $B$ located on the destination's trajectory (before the rendez-vous point $A$).
We also take $B$ such that the distance from the rendez-vous point $A$ is $r_B=\Theta(\sqrt{r})$.
In the first stage, the information is transmitted to new nodes (according to the above angle restriction and ensuring a journey capacity at least $y$) until reaching a node, whose trajectory's distance from $B$ is at most $\sqrt{r}$.

In the second stage the node with the information simply travels a straight line (of length $r+O(\sqrt{r})$)
until approaching the point $B$ within distance $\sqrt{r}$.

In the third stage, the information is transmitted to new nodes (again, with a relative direction angle in $[\frac{a}{2},a]$, and ensuring a journey capacity at least $y$) until
the information is transmitted to a node that passes within distance $1$ of the rendez-vous point $A$, while the contact duration with the destination is sufficient to transfer all the information.

We will show that this routing scheme guarantees that the information will reach the destination with a journey of capacity at least $y$, with a total journey duration of $\frac{r}{v}+O(\frac{\sqrt{r}}{v})$.
More precisely, we show that the duration of the first and third stages is $O(\frac{\sqrt{r}}{v})$.
Since the duration of the second stage is $\frac{r}{v}+O(\frac{\sqrt{r}}{v})$, a lower bound on the information propagation speed is $v$.

We now analyze the duration of the three routing stages.

\subsubsection{Stage 1}

We introduce the following notations.
Let $C$ be the node that most recently received all the information, moving at speed $\bv_C$.
We define $\phi_C$ as the angle formed between the vector $CB$ (defined by the locations of the node $C$ and the
point $B$) and the speed $\bv_C$, as depicted in Figure~\ref{meetingpoint}.

\begin{lemma}
The duration $t_1$ of stage 1 of the routing scheme is $\Theta(\frac{\sqrt{r}}{v})$, almost surely.
The distance traveled is $O(\sqrt{r})$.
\label{lem:stage1}
\end{lemma}

\begin{proof}
See appendix.
\end{proof}

\subsubsection{Stage 2}
\begin{lemma}
The duration $t_2$ of stage 2 of the routing scheme is $\frac{r}{v}+O(\frac{\sqrt{r}}{v})$, almost surely.
\label{lem:stage2}
\end{lemma}

\begin{proof}
The initial distance $SB$ is at most $r+r_B=r+O(\sqrt{r})$.
From Lemma~\ref{lem:stage1}, the distance $r_1=CA$ at the end of stage~1 is $r+O(\sqrt{r})$.
The minimum distance of node $C$ trajectory to $B$, and is at most $r_2=r_1 \sin (\frac{1}{\sqrt{r}})=\sqrt{r}+O(r^{-\frac{1}{2}})$, as depicted in Figure~\ref{lowerbound}.
Therefore, there is a point in the trajectory such that the final distance of node $C$ from the point $B$ is exactly $\sqrt{r}$.
Therefore, the total distance traveled in stage~2 is at most $r_1 (1+(\frac{1}{\sqrt{r}})) = r+O(\sqrt{r})$.
\end{proof}

\subsubsection{Stage 3}
At the beginning of stage~3, there is a node carrying the information, located within distance $r_B+\sqrt{r}$ from the rendez-vous point, and within distance $\sqrt{r}$ from the destination's trajectory.
In this stage, the information is transmitted to new nodes (again, according to the above angle restriction and ensuring a capacity at least $y$) until
reaching a node that passes within distance $1$ of the rendez-vous point $A$, while the contact duration with the destination is at least $y$.

Equivalently to stage~1, let $C$ be the node that most recently received all the information, moving at speed $\bv_C$.
We introduce again the angle $\phi_C$, this time defined with respect to the rendez-vous point $A$; namely, $\phi_C$ is the angle formed between the vector $CA$ (defined by the locations of the node $C$ and the
rendez-vous
point $A$) and the speed $\bv_C$.

\begin{lemma}
We consider a node $C$, at distance $r_C$ from the rendez-vous point, moving with speed $\bv_C$ at a direction such that the relative angle with the destination's direction is at most $a=\frac{1}{2uy}$.
If the angle $\phi_C$
is at most $\frac{1}{2 r_C}$, then the trajectory of $C$ passes within range of the destination and guarantees that the meeting duration with a destination located at $A$, moving at constant speed, will be at least equal to~$y$.
\end{lemma}

\begin{proof}
The relative speed of the node $C$, with respect to the destination's speed, is at most $2v\sin (\frac{a}{2}) \leq v a$.
If the node $C$ passes within distance $m$ from the rendez-vous point, the meeting duration is at least $\frac{1-m}{va}$ (since the distance traveled within range, in the frame of reference of the destination, is at least $1-m$).
Therefore, in order for the meeting duration $T$ to be at least equal to $y$, it is sufficient that:
$
m \leq 1- y va = \frac{1}{2}
$.
In this case, we guarantee a meeting duration at least equal to $y$.
Moreover, if we have $\phi_C \leq \frac{1}{2 r_C}$, the node will pass within distance $\frac{1}{2}$ from the rendez-vous point.
\end{proof}

\begin{lemma}
The duration $t_3$ of stage 3
is $O(\frac{\sqrt{r}}{v})$, almost surely.
At the end of stage~3, the destination is reached at the rendez-vous point with probability strictly larger than $0$.
\label{lem:stage3}
\end{lemma}

\begin{proof}
See appendix.
\end{proof}

\begin{theorem}
Consider a network with constant node density $\nu$, radio range $R$ and communication rate $G$, where nodes move at speed $v>0$ and change direction at rate $\tau=0$.
When the journey
capacity is at most $y=\frac{K}{v}$, where $K$ is a constant, a lower bound on the information propagation speed is $s_L(y)=v$.
\label{theo:lower}
\end{theorem}

\begin{proof}
Considering the final position of any destination, we can define a rendez-vous point $A$.
If the distance of the rendez-vous point from the source location at time $0$ is $r \to \infty$, based on the previous lemmas, there exists with strictly positive probability a journey of capacity at least $y$ that reaches any rendez-vous point $A$ within time $\sim \frac{r}{v}$.
Therefore, the asymptotic information speed is at least $v$.
\end{proof}

We note that, in case the network domain $\A=L\times L$
is sufficiently large,
for all destination nodes which, at time $t=\Theta(L)$, are at distance $r=o(v t)$ of the initial source location, there is almost surely a journey of duration $t$ and of capacity $y$ from the source to the destination.

\begin{remark}
Although, we derived the lower bound in a pure billiard mobility model, the proof can be easily generalized to a random walk model, where the change of direction rate is $O(\frac{1}{r})$, by restarting from the first stage at any change of direction (an event which occurs a finite number of times).
\end{remark}

\section{Upper Bound and Space-Time Capacity}
\label{Sect:upper}

In this section,
our aim is to find the shortest journey of capacity at least $y$
that
connects any source to any destination in the network domain.
We prove an upper bound $s_U(y)$
on the information propagation speed, for
journeys of capacity~$y$.

\begin{theorem}
Consider a network with $n$ mobile nodes with radio range~$R$, communication rate $G$, in a square area of size $\A=L\times L$, where nodes move at speed $v$, and change direction at rate~$\tau$. When $n\to\infty$, such that the node density becomes $\nu=\frac{n}{L^2}$,
an upper bound on the information propagation speed, for journeys of capacity $y$,
is the smallest ratio of $\frac{\theta}{\rho}$ with:
{\small
$$
\min_{\rho,\theta>0}\left\{\frac{\theta}{\rho}~\text{with}~
\theta=\sqrt{\rho^2 v^2+\left(\tau+\frac{\gamma(y) 4 \pi v \nu R I_0(\rho R)}{1-\gamma(y) \frac{\pi \nu R}{2 \rho}I_1(\rho R)}\right)^2}-\tau\right\},
$$}
where
$I_0()$ and $I_1()$ are {\em modified Bessel functions},
and,
\begin{itemize}
\item $\gamma(y)=\min(\frac{\pi^2 R G}{8 v y},1)$, if $\tau > 0$;
\item $\gamma(y)=\min(\frac{(R G)^2}{3(vy)^2},1)$, if $\tau =0$.
\end{itemize}
\label{theo:upper}
\end{theorem}

\begin{remark}
The expression of $\theta$ has meaning
when $\pi \nu R^2 \gamma(y)<1$.
Above this threshold,
the upper bound for the information propagation speed is infinite.
Such a behavior is expected, since 
there exists a critical node density above which the graph is fully connected or at least
percolates~\cite{MR96}.
In addition,
according to Theorem~\ref{theo:upper},
in percolated networks, there is a critical journey capacity $y_c$, such that,
when $y>y_c$, the propagation speed is bounded by a constant.
\end{remark}

\begin{proof}
We assume that a source starts emitting information at position $\bz=0$ and time~$t=0$.
We consider the probabilistic space-time journey model presented in Section~\ref{Sect:thickness}, which includes
all shortest journeys originating at the source.
Equivalently, we model journeys of very small beacons of information, such that beacon transmissions are instantaneous.

We initially consider an infinite network with a Poisson density of nodes $\lambda$.
We will upper bound the probability density of journeys in the infinite network model.
However,
by applying an analytical depoissonization technique~\cite{js98},
we obtain an equivalent asymptotic estimate of the journey density when the number of nodes $n$ is large but not random.

We decompose the journeys into two types of segments, modeling node movements and beacon transmissions:
\begin{itemize}
\item \emph{emission segments} $S_e(\bu,\bv)$: the node
transmits immediately after receiving the beacon;
$\bv$ is the speed of the node that just received the beacon, and $\bu$ is the
emission space vector and is such that $|\bu|\le R$;
\item \emph{move-and-emit segments} $S_m(\bu,\bv,\bw)=M(\bv,\bw)+\bu$: $M(\bv,\bw)$ is the
space-time vector corresponding to the motion of the node carrying the beacon,
where $\bv$ is the initial
vector speed of the node when it receives the beacon and
$\bw$ is the final speed of the node just
before transmitting the beacon; the vector $\bu$ is the emission space vector
which ends the segment.
\end{itemize}

Considering any sequence of segments,
we can always upper bound the segment probabilities (see~\cite{JMR09}, Section III-B).
In fact, the conditional probabilities, given the node direction and speed, are upper bounded by unconditional probabilities:
\begin{itemize}
\item $\TP(S_e(\bu))=P(\bu) \lambda$,
where $P(\bu)$ is the probability density of $\bu$ inside the disk of radius~$R$,
and~$\lambda$ is the node density
(to make the emission possible);
\item $\TP(S_m(\bu,\bv,\bw))=P(\bu)P(M(\bv,\bw)) 2v \lambda$, where $P(\bu)$ is the probability density of $\bu$ on the circle of radius~$R$ (we only need to consider the earliest transmissions, which occur at the maximum radio range), $P(M(\bv,\bw))$ is the probability that the node movement equals the space vector $M(\bv,\bw)$, and
$v$ is the node speed.
\end{itemize}
This upper bound journey model results in a higher density of journeys than in the actual network.
But, in this model,
any journey can be decomposed into a sequence of independent segments.
Consequently,
we can express a journey $\CC$ as an arbitrary sequence of emission \emph{or} move-and-emit segments, \ie using regular expression notation, $\CC=(S_e+S_m)^*$.
Moreover, we can calculate the Laplace transform of the journey probability density, based on the Laplace transforms of the segments.
We denote the segment Laplace transforms by $l_e(\zeta,\theta)=E(e^{-(\zeta,\theta) \cdot S_e})$ and $l_m(\zeta,\theta)=E(e^{-(\zeta,\theta) \cdot S_m})$, for emission and move-and-emit segments, respectively.
Equivalently to the formal identity $\frac{1}{1-x}=1+x+x^2+x^3+...$, which represents the Laplace transform of an arbitrary sequence of random variables with Laplace transform $x$, the journey Laplace transform
has a denominator $k(\zeta,\theta)$, equal to:
\begin{equation}
k(\zeta,\theta)=1-\left(l_e(\zeta,\theta)+l_m(\zeta,\theta)\right).
\label{eq:kernel}
\end{equation}

We have the following Laplace transform expressions:
\begin{itemize}
\item $l_e(\zeta,\theta)=E(e^{-\zeta \cdot \bu})$, where $\bu$ is uniform in the disk of radius $R$, with density $\lambda$,
\ie $l_e(\zeta,\theta)=\lambda \pi \frac{2 R}{|\zeta|}I_1(|\zeta| R)$.
\item $l_e(\zeta,\theta)=E(e^{-\zeta \cdot \bu}) E(e^{-(\zeta,\theta) \cdot M(\bv,\bw)})$, where $\bu$ is uniform on the circle of radius $R$, with density $\lambda$,
\ie $E(e^{-\zeta \cdot \bu}) = 2 \pi \lambda R I_0(|\zeta| R)$, and $E(e^{-\sigma \cdot M(\bv,\bw)})=\frac{1}{\sqrt{(\theta+\tau)^2-|\zeta|^2v^2}-\tau}$ (see~\cite{JMR09}).
\end{itemize}

We derive an upper bound on the information propagation speed,
in the special case where the journey capacity is $y=0$,
from the analysis of the singularities of the
journey Laplace
transform, for $\lambda$ equal to the node density in the network ({\it cf.} Theorem~1 in~\cite{JMR09}).
The upper bound is the smallest ratio $\frac{\theta}{\rho}$ of the non-negative pair $(\rho,\theta)$ which is a root of the denominator~$k(\rho,\theta)$ (with $\rho=|\zeta|$), obtained by substituting the segment Laplace transforms expressions in~(\ref{eq:kernel}).

In order to generalize to journeys of a given capacity $y>0$, we will restrict the set of possible journeys, to those satisfying the desired capacity constraint, and calculate the Laplace transform of the journey density in this restricted set.

First, we remark that, a journey has a capacity at least~$y$, if and only if the journey \emph{thickness} (\ie the minimum duration of all data transmissions in the journey) is at least equal to~$\frac{y}{G}$, with $G$ the communication rate.
Therefore, when considering journeys of a given capacity, we can equivalently focus on the possible journeys with minimum node meeting duration~$\frac{y}{G}$.

Therefore,
in the upper-bound journey model, we can substitute the probability of any emission segment with
the probability of the same emission segment, while additionally ensuring that the emission duration is at least~$\frac{y}{G}$.
Thus, for the singularity analysis, we substitute in~(\ref{eq:kernel})
the Poisson density $\lambda$ with a node density~$\nu \gamma(y)$,
where~$\gamma(y)$ is an upper bound on the probability that the meeting duration is at least~$\frac{y}{G}$.
This direct substitution is feasible because we work with the upper bound journey model, where 
successive segments (including all transmissions) are independent of the previous network state.
Again, this results in considering a higher density of journeys than in the actual network (including some transmissions which are not actually possible,
due to the node directions); however, there is no impact on the validity of our analysis, since we are interested in upper bounds.

To conclude the proof, it suffices to substitute quantity~$\gamma(y)$ using Lemma~\ref{lem:distT} when $\tau=0$, and Lemma~\ref{lem:meetingbound} when~$\tau>0$.
\end{proof}

We derive the following corollaries expressing the behavior of the upper bound when the journey capacity is large, in random waypoint-like ($\tau \to 0$) and random walk/Brownian motion mobility ($\tau>0$), respectively.

\begin{corollary}
When nodes move at speed $v>0$, and $\frac{\nu}{y}\to 0$
(\ie the journey capacity $y$ is large)
with $\tau > 0$,
the propagation speed upper bound is
$O(\sqrt{\frac{\nu G}{y \tau}}R v)$.
\label{cor:randomwalk}
\end{corollary}
\begin{proof}
See appendix.
\end{proof}

\begin{corollary}
When the node speed is $v>0$, and $\frac{\nu}{y}\to 0$
(\ie the journey capacity $y$ is large)
with
$\tau = O(\frac{1}{L}) \to 0$,
the propagation speed upper bound is
$v+O(\tau +\frac{\nu G R^2}{y})$.
\label{cor:billiard}
\end{corollary}
\begin{proof}
See appendix.
\end{proof}

We observe that, for large journey capacities~$y$,
the upper bound on the information propagation speed $s_U(y)$ tends to the actual mobile node speed~$v$ in random way-point-like mobility,
while it decreases with the inverse square root of the journey capacity~$y$
in random walk or Brownian motion mobility.
In both cases, the resulting upper bound on the space-time capacity $c(y)=s_U(y) y$ is a function which increases with~$y$.

\begin{remark}
When nodes move at speed $v>0$ in random way-point-like mobility:
\begin{itemize}
\item from Theorem~\ref{theo:lower}, a lower bound on the propagation speed is $v$, for
any bounded $y$, and when the node density is $\nu=\Theta(1)$;
\item from Corollary~\ref{cor:billiard}, an upper bound on the propagation speed is $v$, for journey capacities $y$ such that
$\nu=o(y)$.
\end{itemize}
Therefore, we notice that there is a range of values of $y$, for which our bounds are almost tight.
More generally,
we deduce that the information propagation speed in random way-point-like mobility models is of the same order as the mobile node speed, for (bounded) journey capacities that are relatively large with respect to the node density.

This implies that, in sparse but large-scale mobile DTNs, the space-time information propagation capacity in bit-meters per second remains proportional to the mobile node speed and to the size of the transported data bundles, when the bundles are relatively large. It is rather surprising that the propagation speed does not tend to $0$ when the size of the bundles increases, which would result in a sub-linear increase of the space-time capacity.
\label{rem:bundle}
\end{remark}

\section{Numerical Results}
\label{Sect:simulations}

In this section, we perform simulation measurements to compare to the analytical bounds on the information propagation, derived in the previous sections.
We developed a simulator that follows the network and mobility model described in Section~\ref{Sect:model}.
We simulate the epidemic broadcast of information, and we consider journeys with a given lower bound on the capacity $y$,
as described in Section~\ref{Sect:thickness}.
We note that the simulation is more general than the simple broadcast of a packet of size~$y$,
since the information can also be transferred on a given journey using smaller packets.
In fact,
we precisely ensure that the journeys of the simulated broadcast have a capacity at least~$y$, without imposing further restrictions.
For all the following simulations, we consider a communication rate $G=1$ units of data per second
(\eg if one unit of data corresponds to $x$ Mbits,
the journey capacity in the following examples should be multiplied by~$x$ Mbits).

\begin{figure}[t!]
\begin{center}
\includegraphics[width=5.4cm]{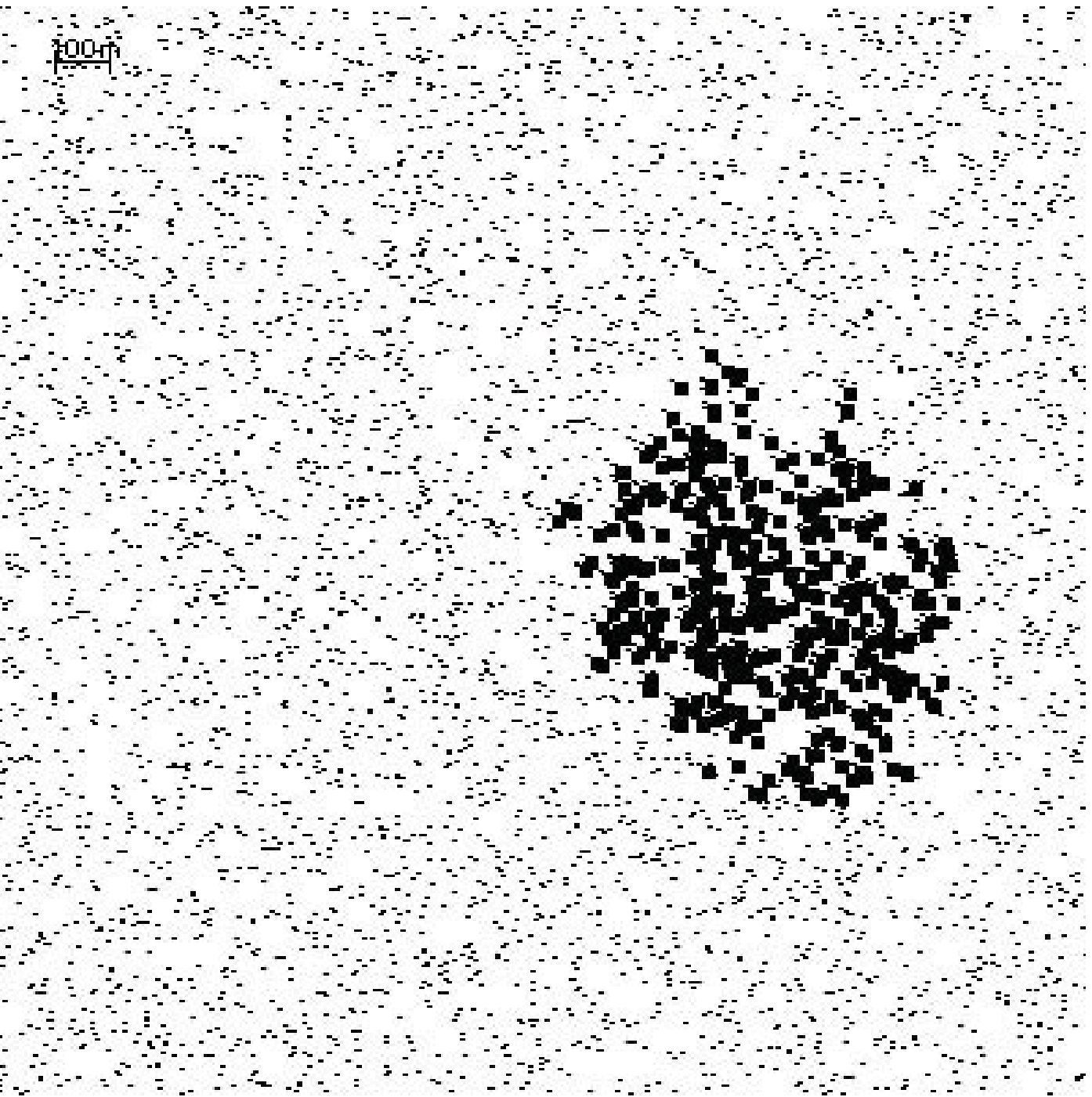}
\hskip 0.3cm
\includegraphics[width=5.4cm]{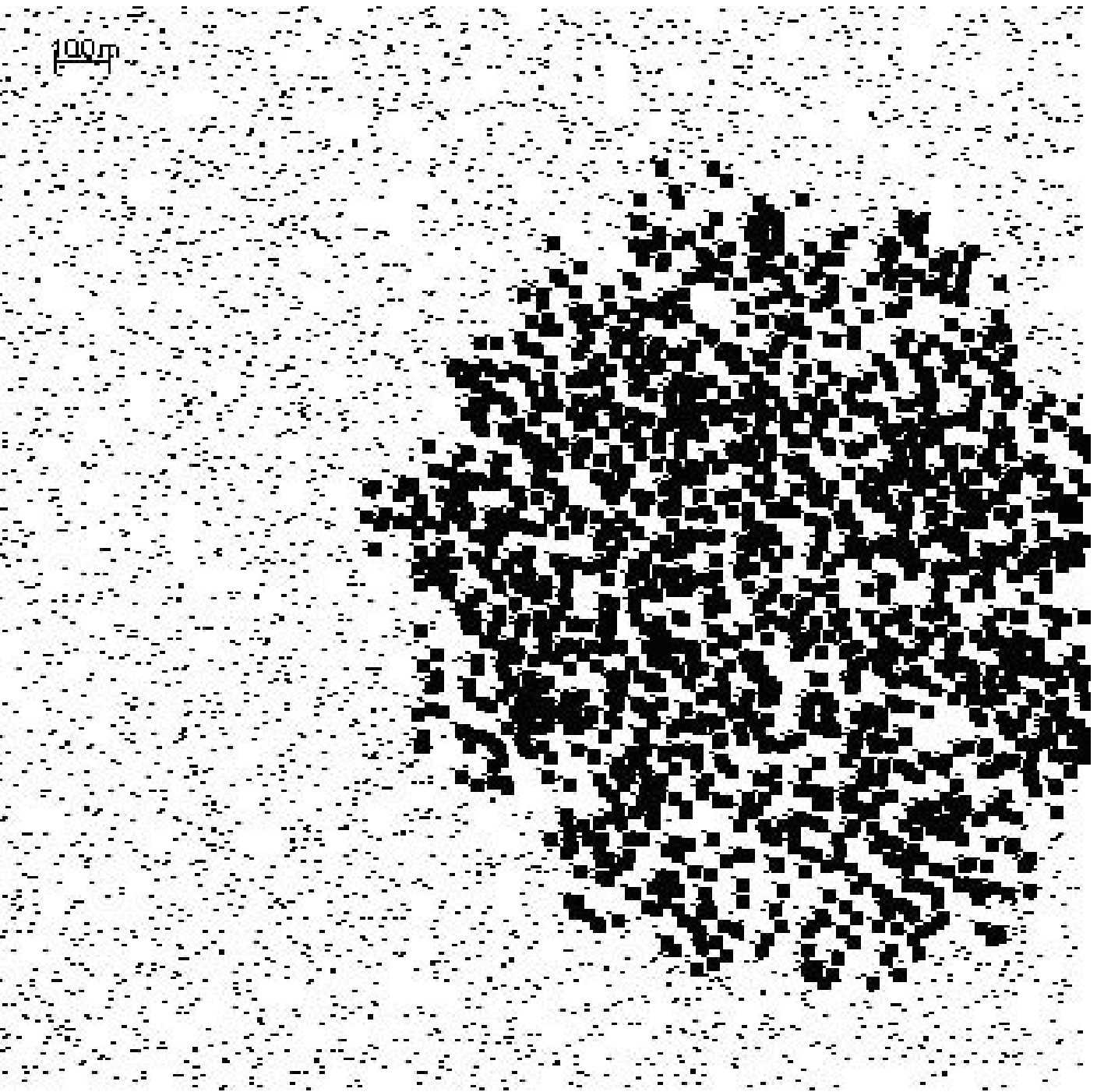}
\hskip 0.3cm
\includegraphics[width=5.4cm]{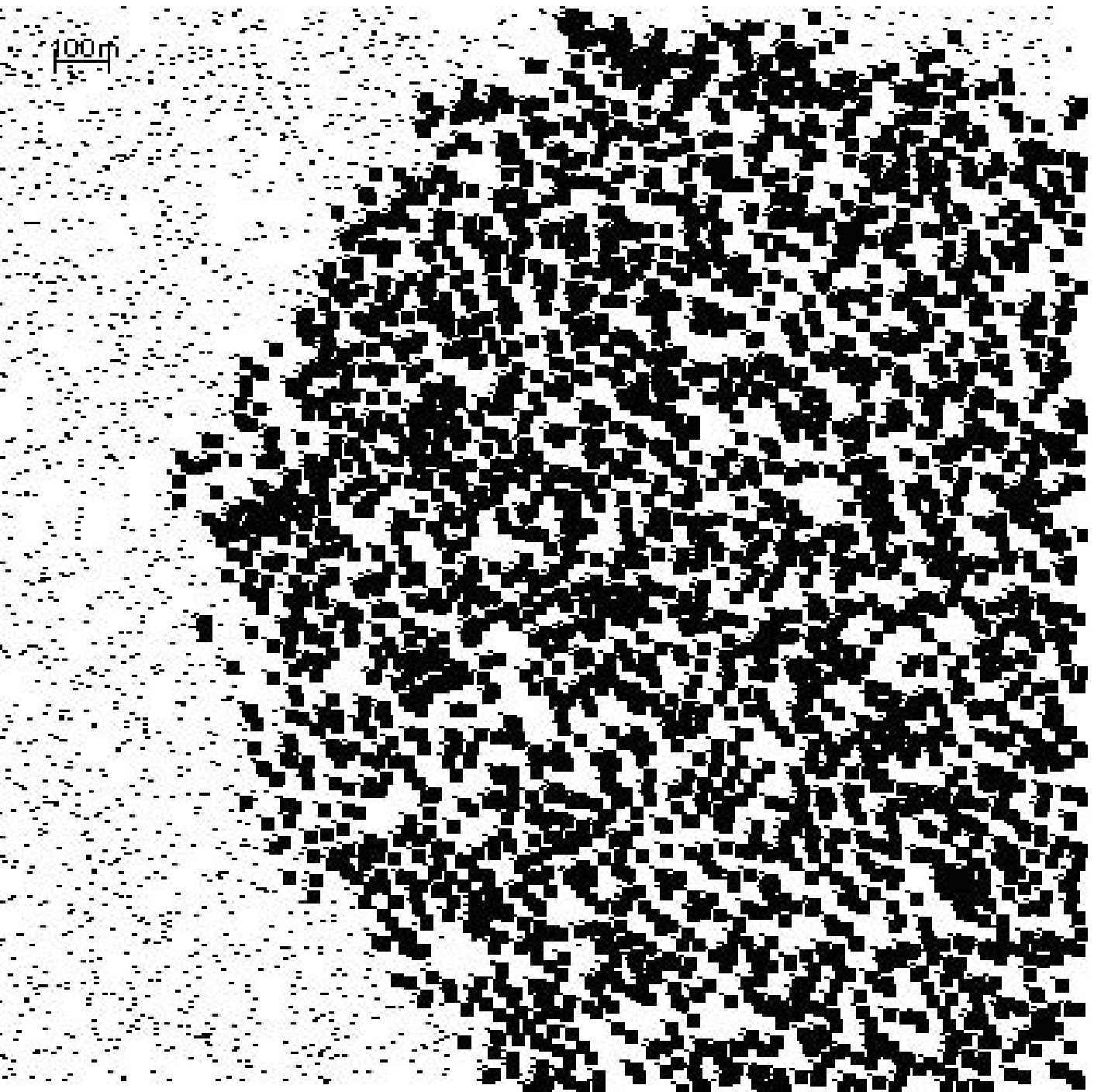}
\vskip 0.3cm
\includegraphics[width=5.4cm]{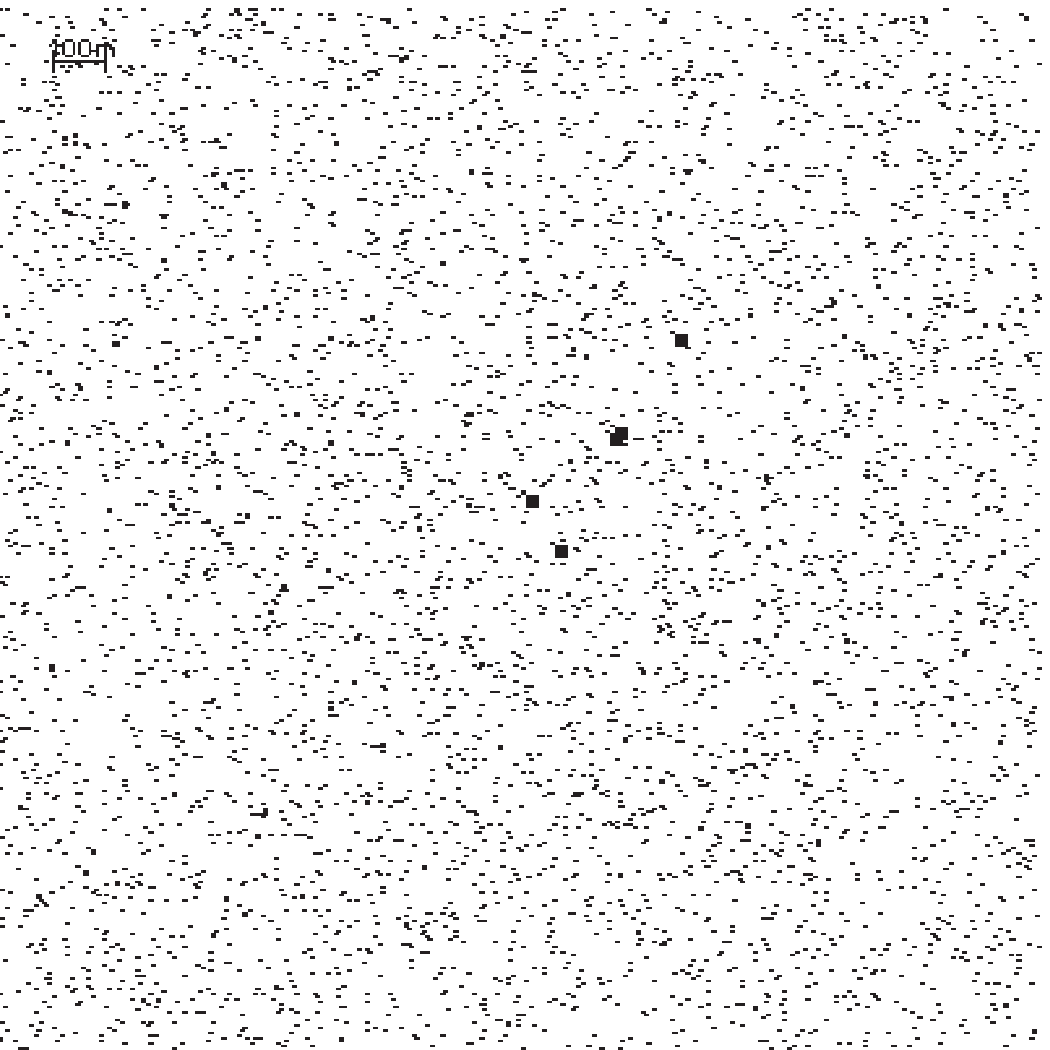}
\hskip 0.3cm
\includegraphics[width=5.4cm]{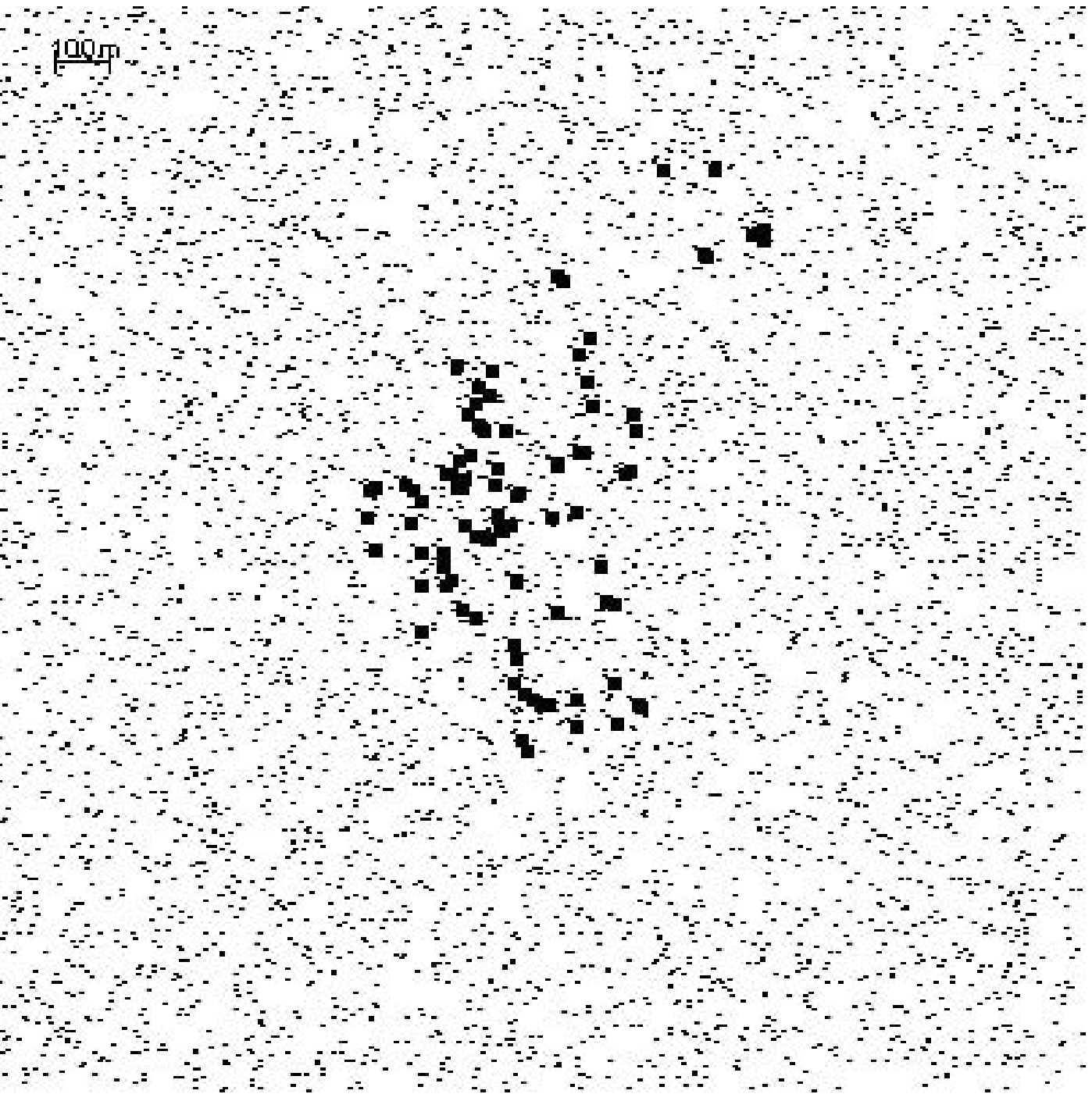}
\hskip 0.3cm
\includegraphics[width=5.4cm]{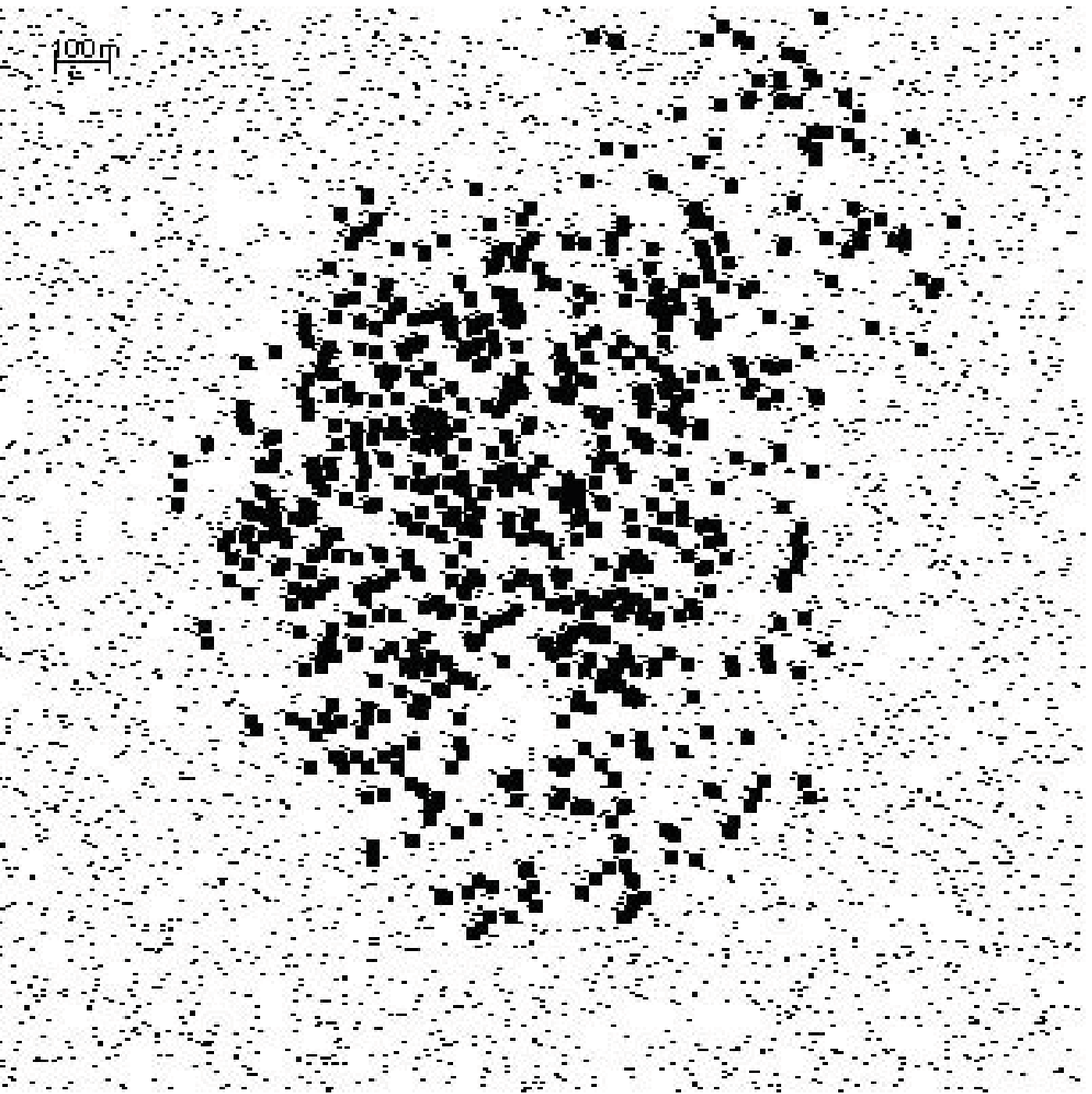}
\caption{Snapshots of simulated information propagation at three different times ($t=100,170,240$), for a small journey capacity $y=0.5$ (top) and a larger journey capacity $y=2.5$ (bottom).
Larger black squares represent nodes that have received all the information at the time of the snapshot.}
\label{fig:propagation}
\end{center}
\end{figure}

We first show how information propagates in a full epidemic broadcast, by illustrating two typical and distinct situations, depending on the journey capacity~$y$.
In the simulated scenario, a source starts broadcasting information at time $t=50$, in a network of $5000$ nodes, in a $2000m \times 2000m$ square, with radio range $R=10m$, and mobile node speed $v=5m/s$, with pure billiard mobility ($\tau=0$).
In Figure~\ref{fig:propagation}, we consider two cases: a smaller journey capacity $y=0.5$ (top) and a larger journey capacity $y=2.5$ (bottom).
For each case, we depict three snapshots of the simulated information propagation at three different times, $t=100,170,240$, from left to right.
The small black dots represent the mobile nodes; when two dots are in contact, the corresponding nodes are within communication range. The larger black squares represent nodes that have received all the information at the time of the snapshot, \ie those that can be reached by a journey of capacity $y$.
The simulation scenario is exactly the same in both the top and bottom figures, with the only change concerning the journey capacities. In both cases, the location of the source is approximately located at the center of the disk containing the black squares, at the top left figure.
We observe that, at the top row of Figure~\ref{fig:propagation} corresponding to a small journey capacity, the information propagates as a full disk that grows at a constant rate, which coincides with the information propagation speed;
all nodes inside the disk can be reached by a journey of capacity~$y$, almost surely. Equivalently, this means that the average information propagation delay scales linearly with the distance from the source, and the ratio of the propagation delay over the distance is equal to the inverse of the information propagation speed.
On the other hand, at the bottom row, corresponding to a larger journey capacity, only some of the nodes inside the disk have been reached by a journey of capacity~$y$.
In this case, the average information propagation delay does not necessarily scale linearly with the distance from the source.
However, the information still propagates at a (smaller than before) maximum speed, equal to the rate at which the disk radius grows.

\begin{figure}[t!]
\begin{center}
\hskip -0.2cm
\includegraphics[height=6.4cm,angle=270]{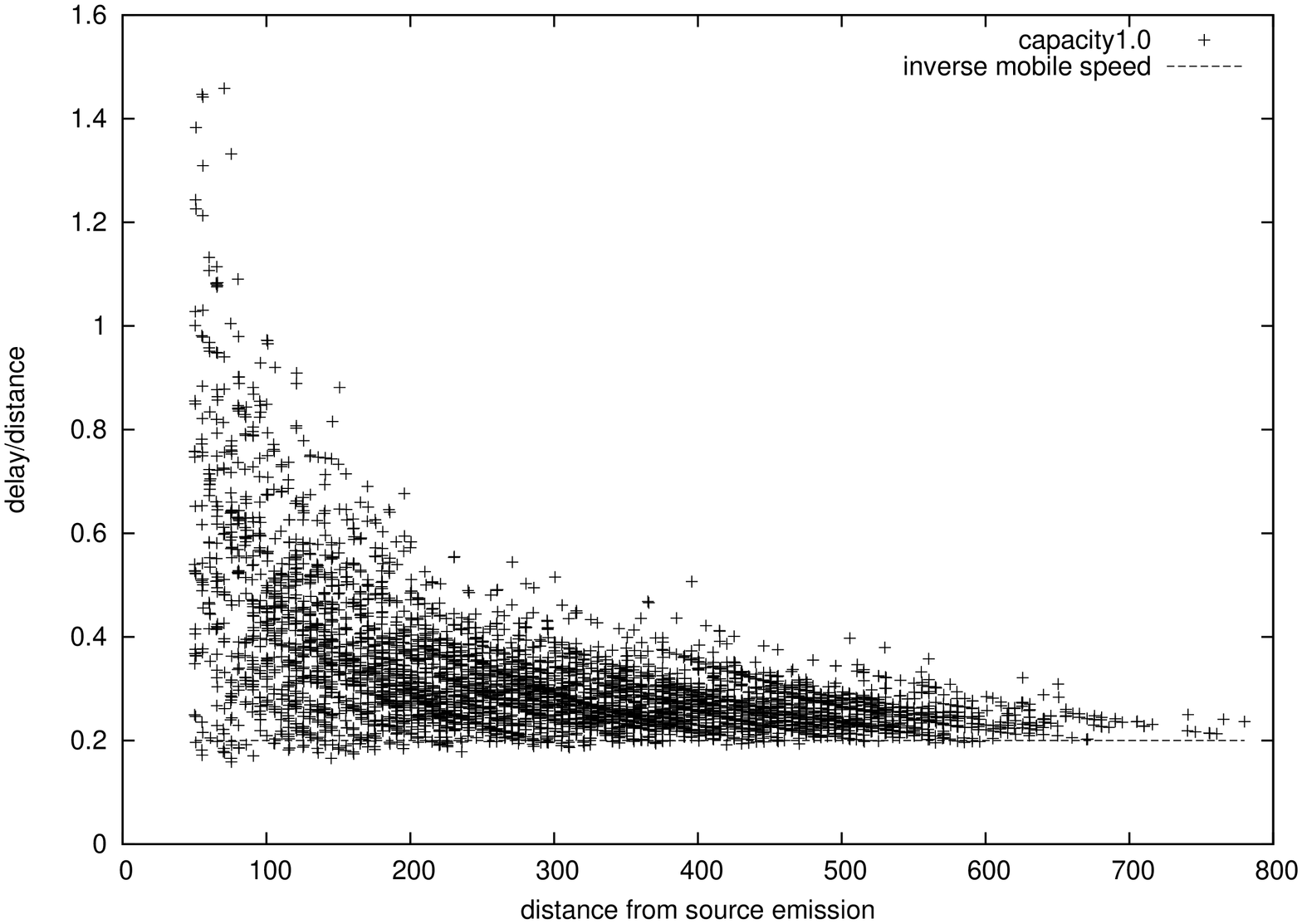}
\hskip -0.cm
\includegraphics[height=6.4cm,angle=270]{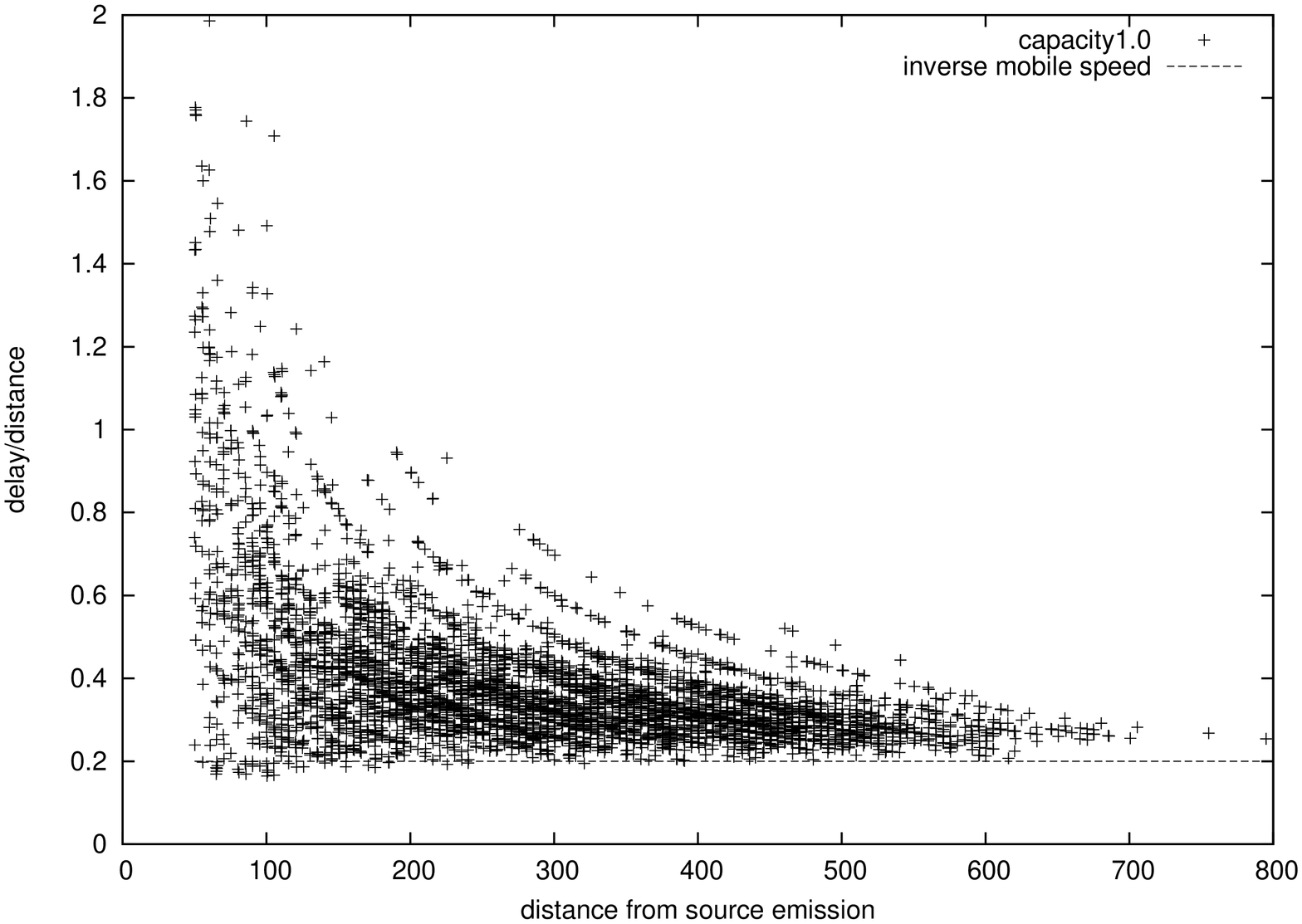}

\hskip -0.2cm
\includegraphics[height=6.4cm,angle=270]{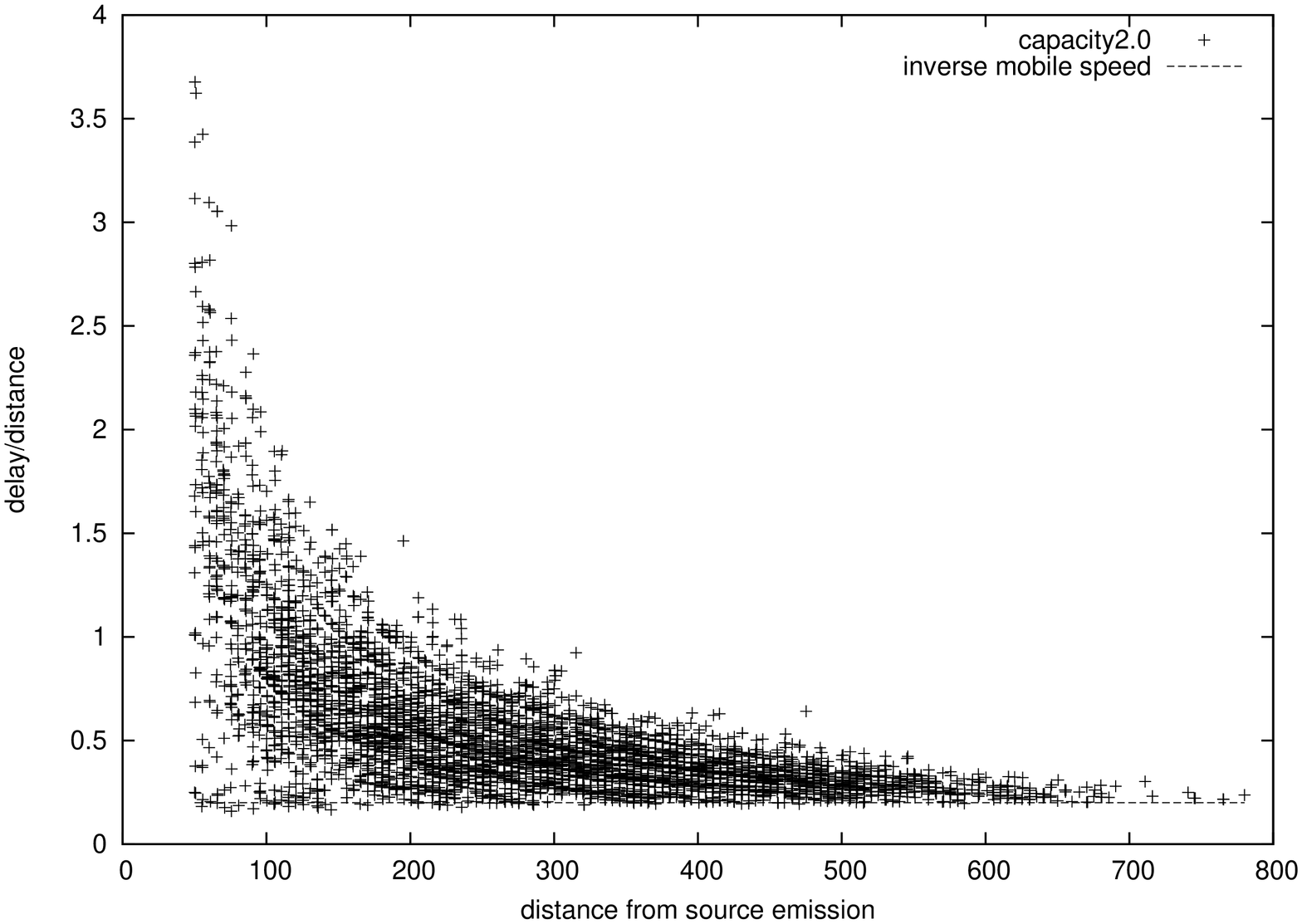}
\hskip -0.cm
\includegraphics[height=6.4cm,angle=270]{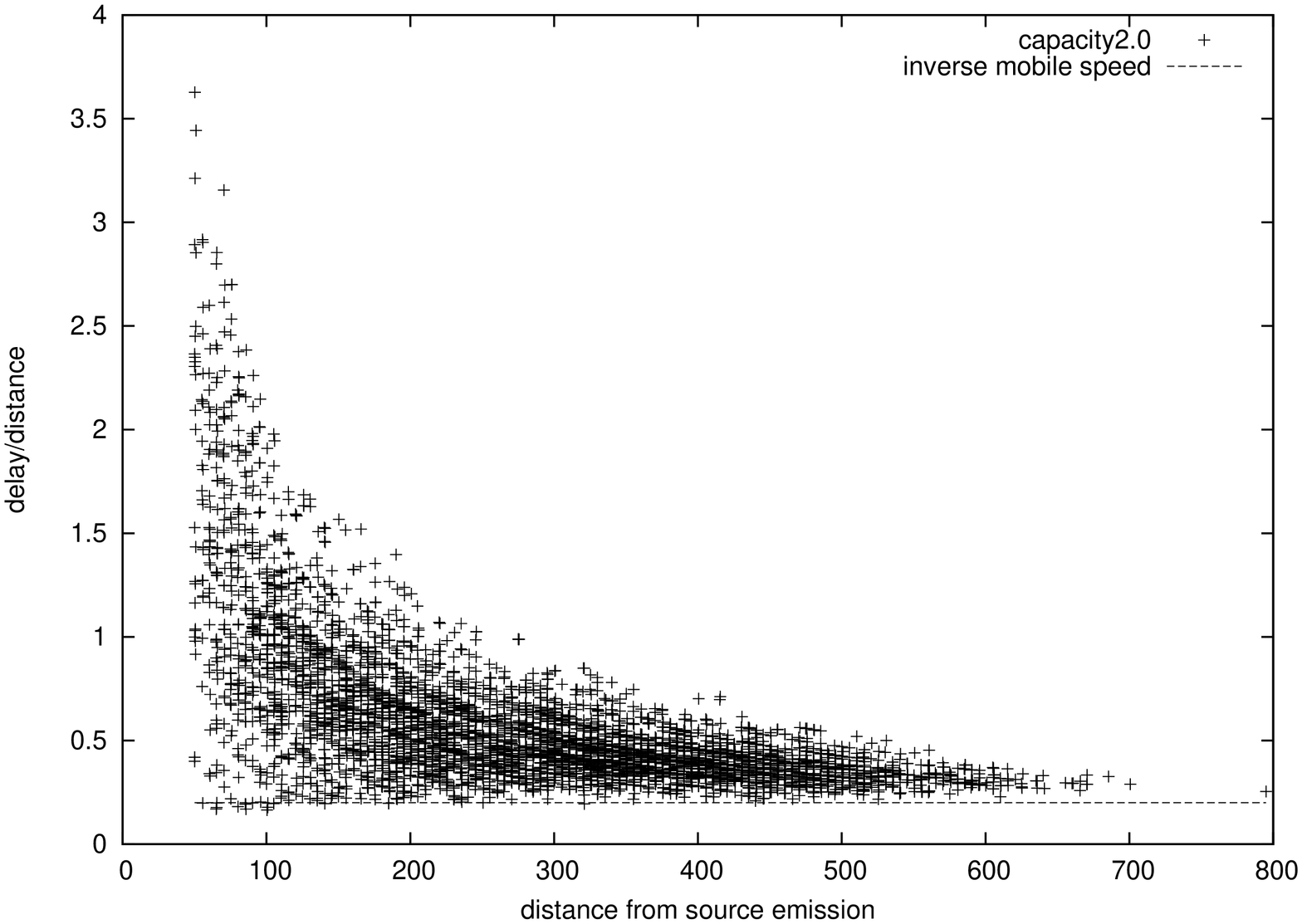}

\hskip -0.2cm
\includegraphics[height=6.4cm,angle=270]{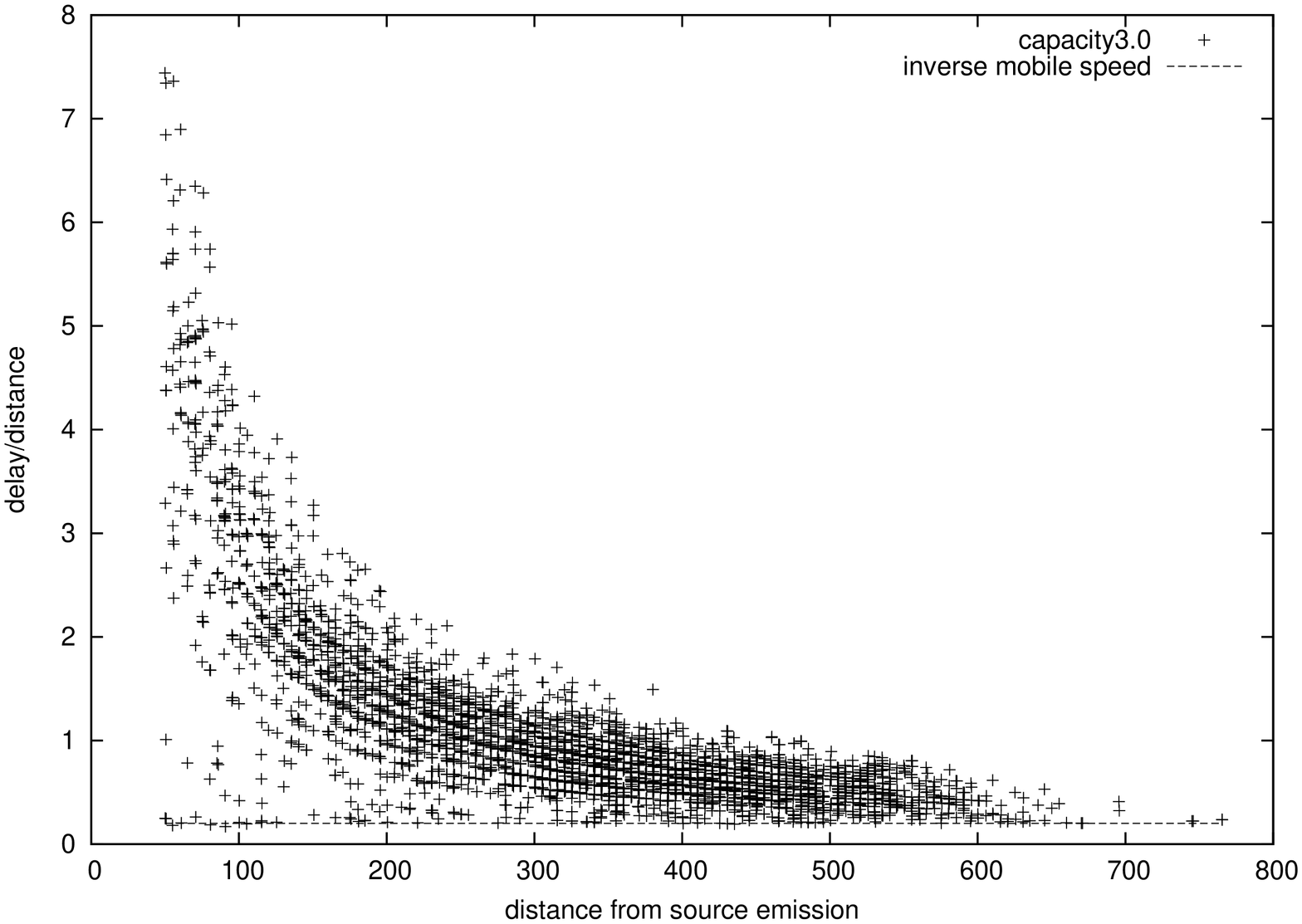}
\hskip -0.cm
\includegraphics[height=6.4cm,angle=270]{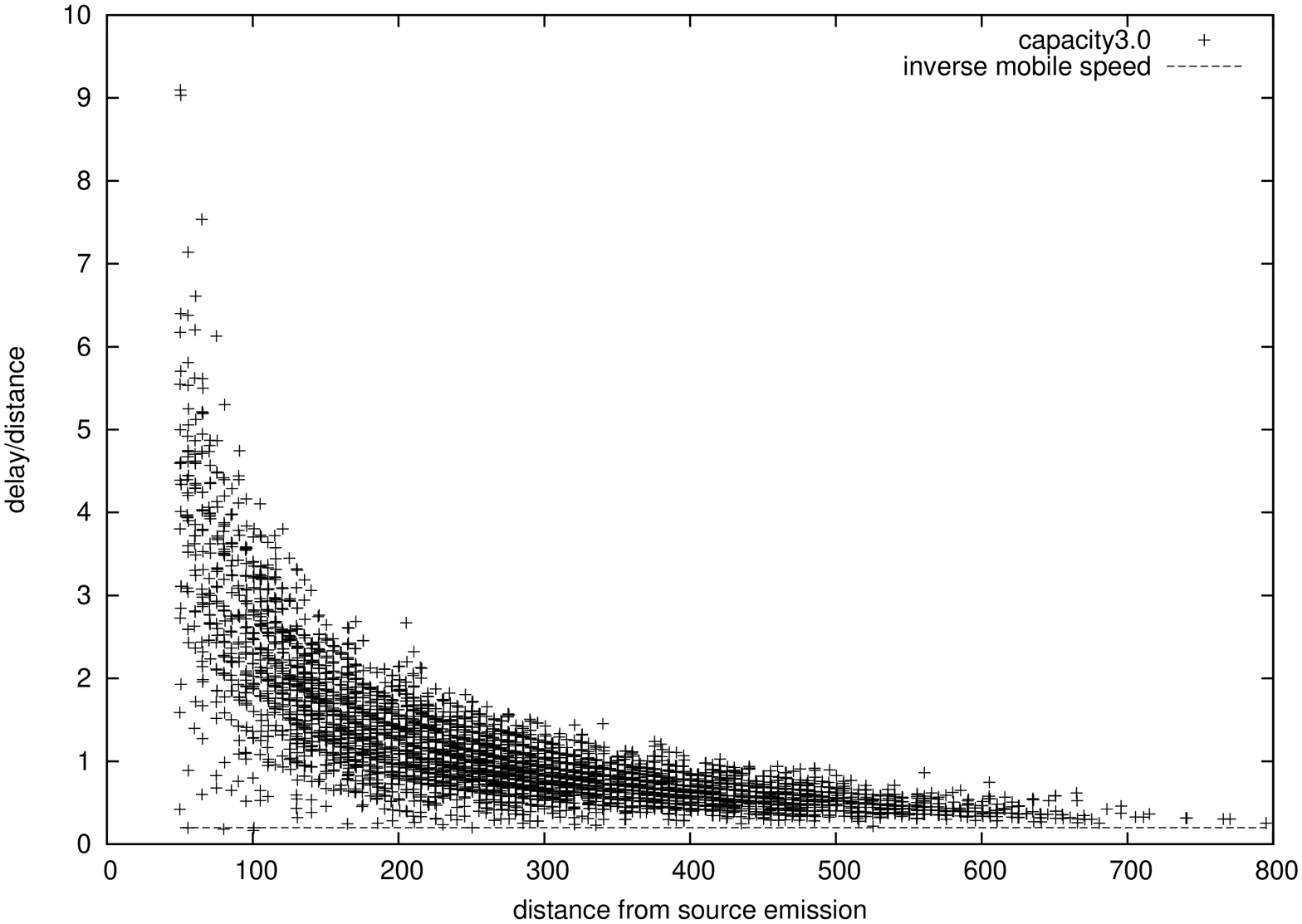}
\caption{Ratio of information propagation delay over distance versus distance from the source, for different journey capacities ($y=\{1;2;3\}$, respectively), compared to the inverse of the mobile node speed, with pure billiard mobility ($\tau=0$ $-$ left), and random walk mobility ($\tau=0.05$ $-$ right).}
\label{fig:delay_vs_distance}
\end{center}
\end{figure}

Next, we simulate a network of $500$ nodes, moving in an area $600m \times 600m$, with a radio range of $10m$, a mobile node speed of $5m/s$ and a communication rate $G=1$ units of data per second.
We simulate two different mobility parameters (rates of direction change): $\tau=0$ for the pure billiard mobility model, 
where nodes change direction only when they bounce on the border, and $\tau=0.05$ for a random walk model. 

In Figure~\ref{fig:delay_vs_distance}
we plot the ratio of the propagation delay over the distance from the source, versus the distance, for journey capacities
$y=\{1;2;3\}$.
Each sample point in the plots corresponds to a simulation measurement.
The distance is measured from the location of the source when the information was emitted to the location of the destination when the information was received.
We notice that, for all journey capacities, the ratio of the propagation delay over the distance is larger than a non-zero constant. The constant lower bound on the ratio, in this simulation scenario, is close to the inverse of the mobile node speed (which is plotted in the figures as a straight line, for comparison).
Furthermore, this constant corresponds to the upper bound on the information propagation speed, which was calculated in Theorem~\ref{theo:upper}.
In fact, for small journey capacities (\eg $y=0.5$), we notice that the upper bound on the information propagation speed is larger than (but close to) the mobile node speed.
For larger journey capacities and $\tau=0$, the upper bound can be obtained from Corollary~\ref{cor:billiard}, and indeed corresponds to the mobile node speed.
We also notice that, for $\tau=0.05$, the average distance that each node travels before changing direction is $100m$, which is of the order of the square network domain length. Therefore, in this case, the upper bound on the propagation speed also remains close to the estimate for random waypoint-like mobility in Corollary~\ref{cor:billiard}, \ie the mobile node speed.

\begin{figure}[t]
\begin{center}
\hskip -0.cm
\includegraphics[height=7.3cm,angle=270]{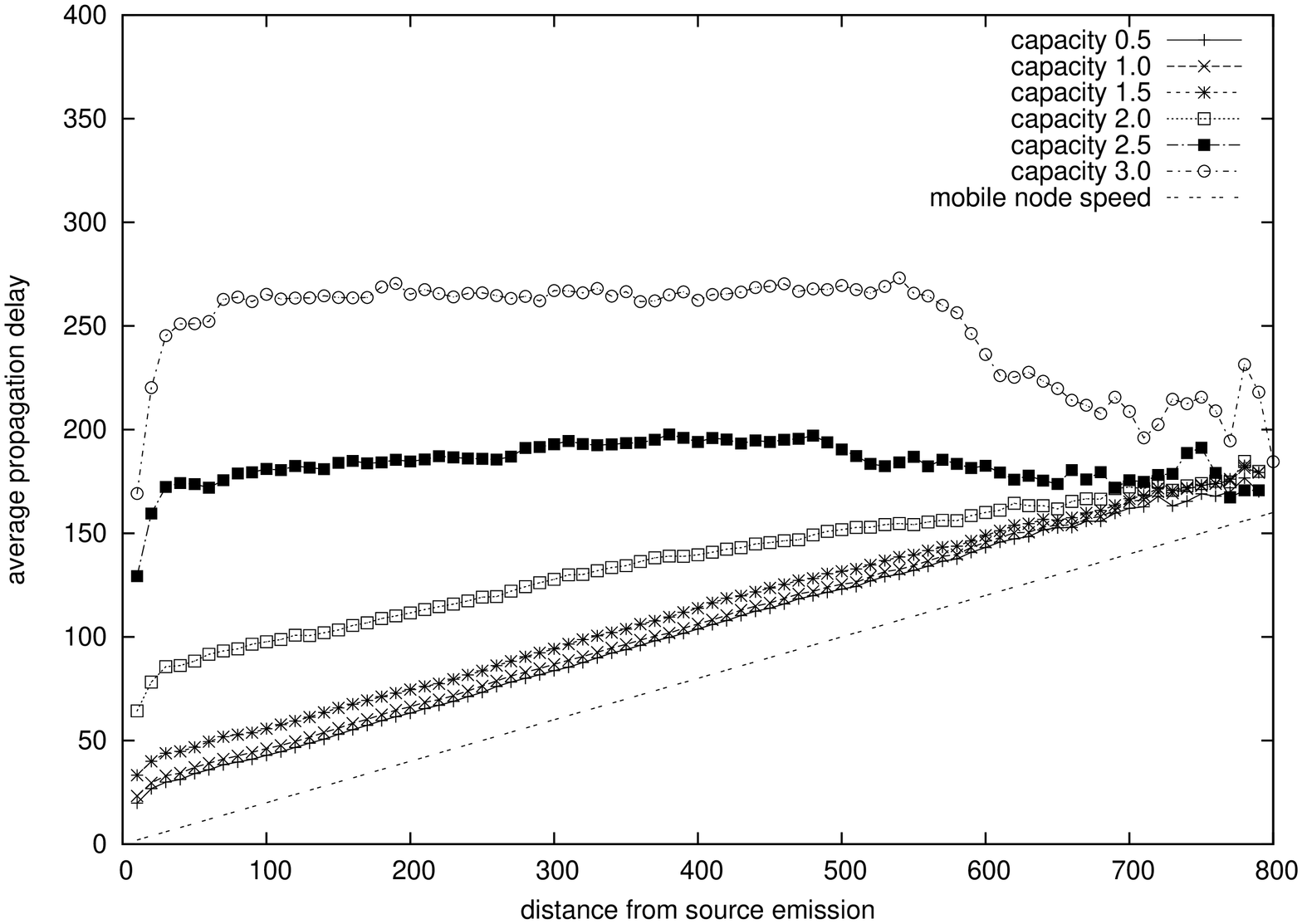}
\includegraphics[height=7.3cm,angle=270]{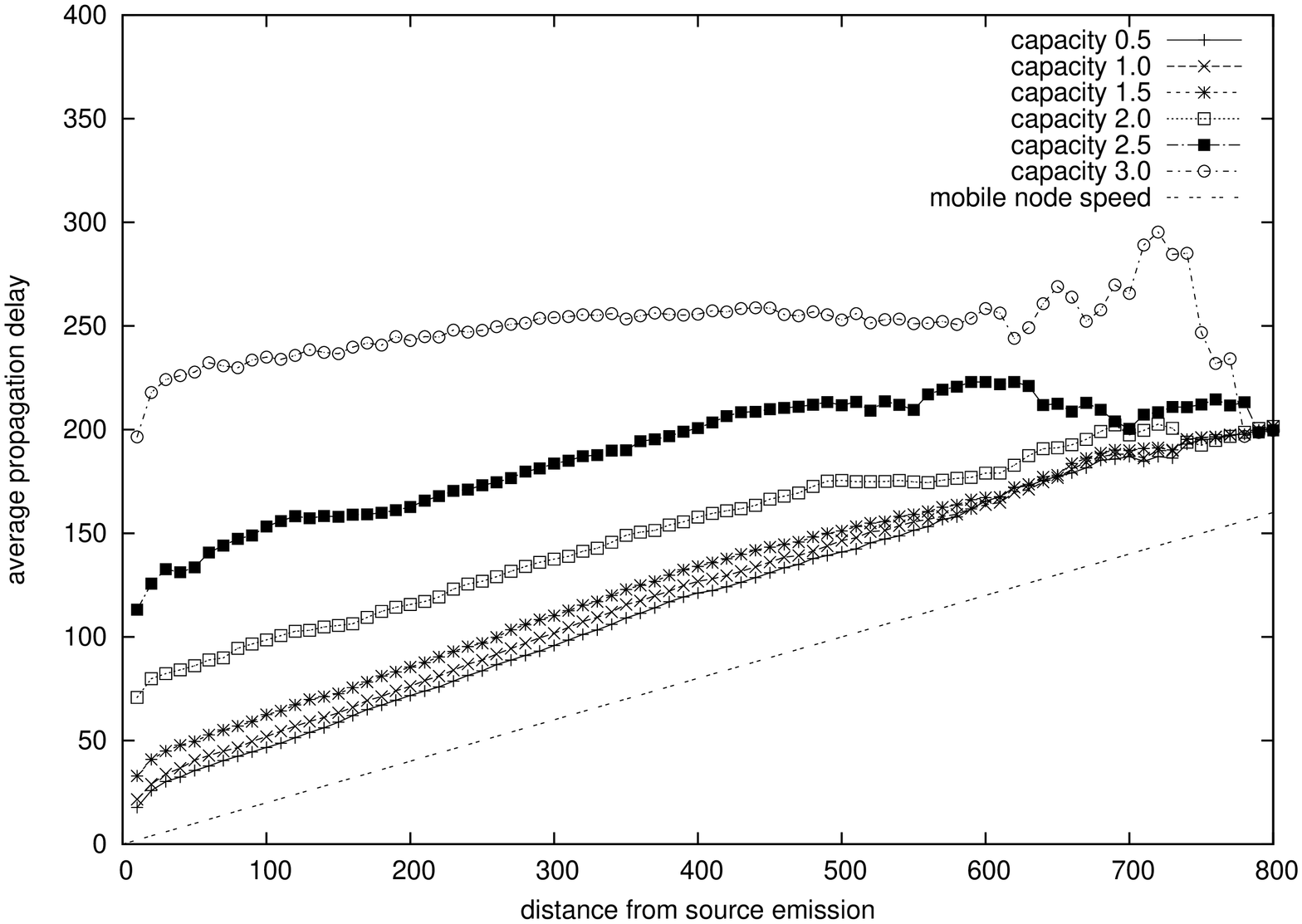}
\caption{Average propagation delay versus distance for different journey capacities ($y=\{0.5;1;1.5;2;2.5;3\}$), with pure billiard mobility ($\tau=0$ $-$ top), and random walk mobility ($\tau=0.05$ $-$ bottom).}
\label{fig:avgdelay_vs_distance}
\end{center}
\end{figure}

\begin{figure}[t]
\begin{center}
\hskip -0cm
\includegraphics[height=6.5cm]{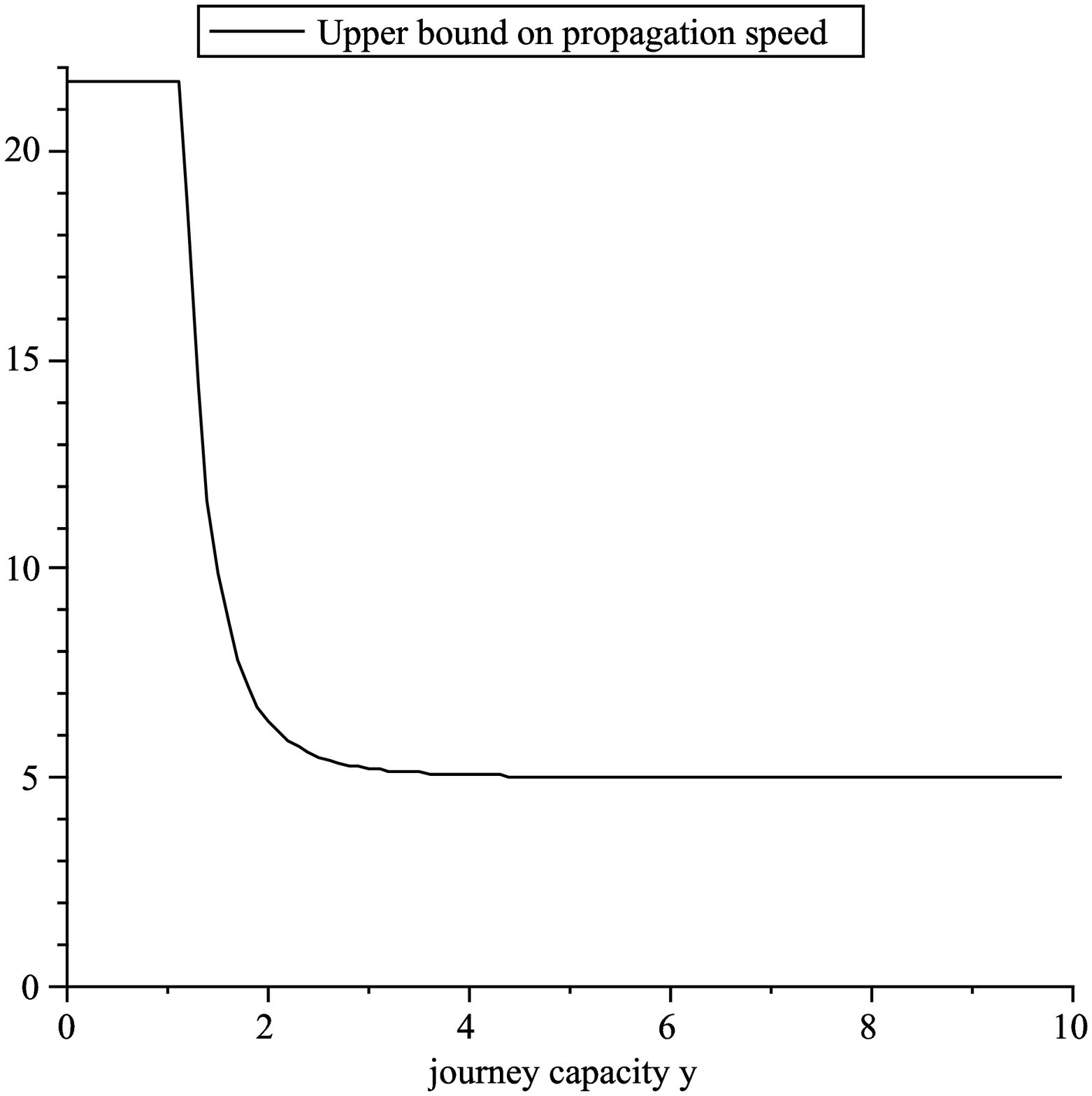}
\hskip 0.2cm
\includegraphics[height=6.5cm]{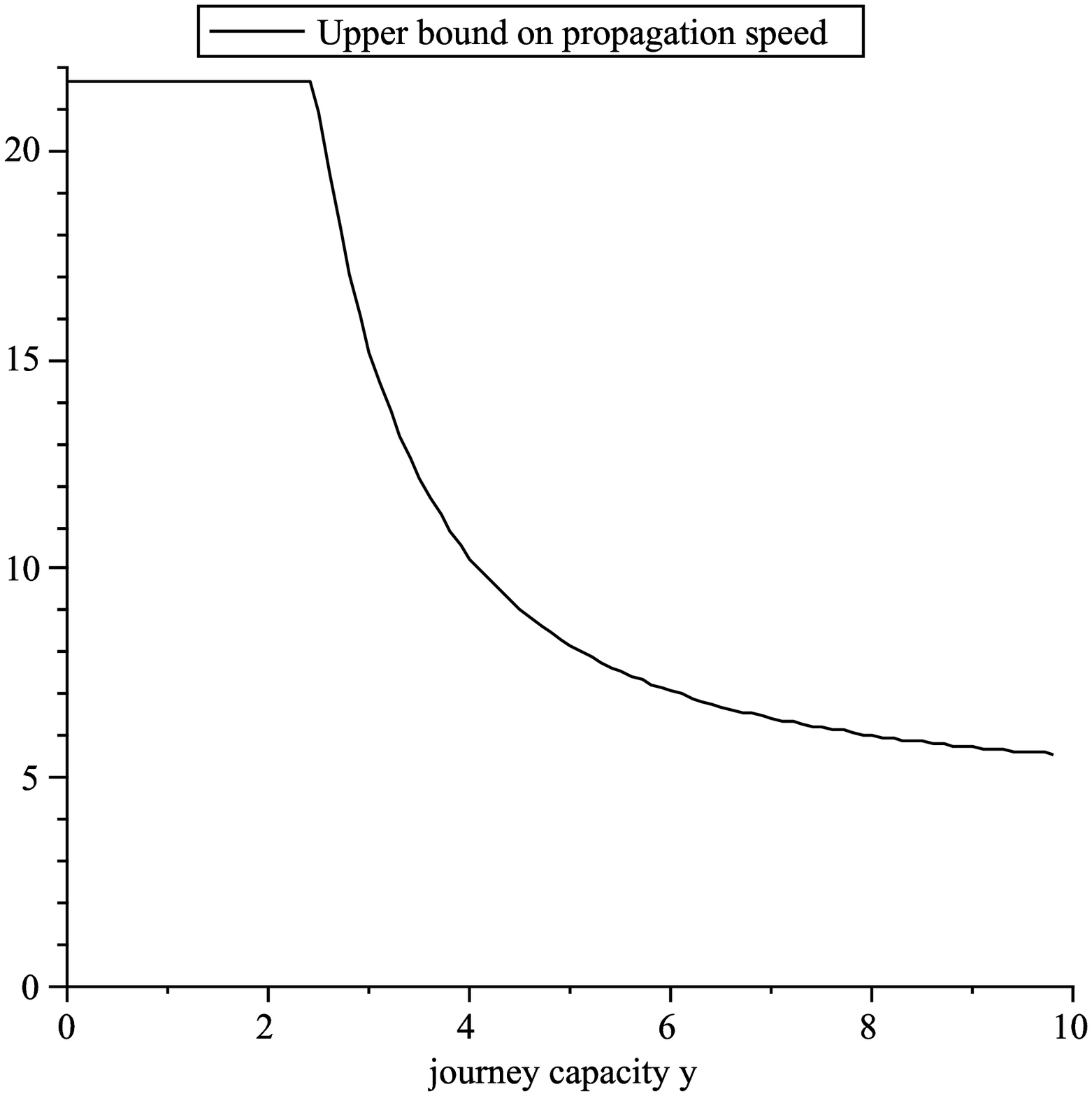}
\caption{Upper bound for the information propagation speed as a function of the journey capacity
($n=500$, $\A=600m \times 600m$, $R=10m$, $v=5m/s$, $G=1$ units of data per second),
with pure billiard mobility ($\tau=0$ $-$ left), and random walk mobility ($\tau=0.05$ $-$ right).}
\label{fig:upperbound}
\end{center}
\end{figure}

In Figure~\ref{fig:avgdelay_vs_distance},
we depict the simulated \emph{average} propagation time versus the distance, for several different journey capacity values $y=\{0.5;1;1.5;2;2.5;3\}$.
Time is measured in seconds, and distance in meters, therefore, the inverse slope of the plots provides us with the information propagation speed in~$ms^{-1}$.
We compare it to a line of fixed slope corresponding to the mobile node speed.
For comparison, we plot the theoretical upper bounds on the information propagation speed (derived from Theorem~\ref{theo:upper}) in Figure~\ref{fig:upperbound}. Simulations show that the theoretical speed is clearly an upper bound.
Moreover, we notice that the upper bound in the case corresponding to random waypoint-like mobility is tighter, due to the fact that our analysis of the node encounter duration analysis (see Lemma~\ref{lem:meetingbound}) is exact in this case.

In Figures~\ref{fig:avgdelay_vs_distance},
we also notice that, for journey capacities up to $2$
units of data per second, the measurements rapidly converge to a straight line of fixed slope, which implies a fixed information propagation speed,
as
illustrated by
the top row of Figure~\ref{fig:propagation}.
However, for larger journey capacities, border effects become significant and the slope of the measurements tends to $0$; this means that, although the maximum information propagation speed is still a non-zero constant, the information does not propagate uniformly as a disk growing at constant speed. In this case, information propagation occurs similarly to the expectation illustrated in the bottom row of Figure~\ref{fig:propagation}.

\begin{figure}[t]
\begin{center}
\hskip -0.cm
\includegraphics[height=7.3cm,angle=270]{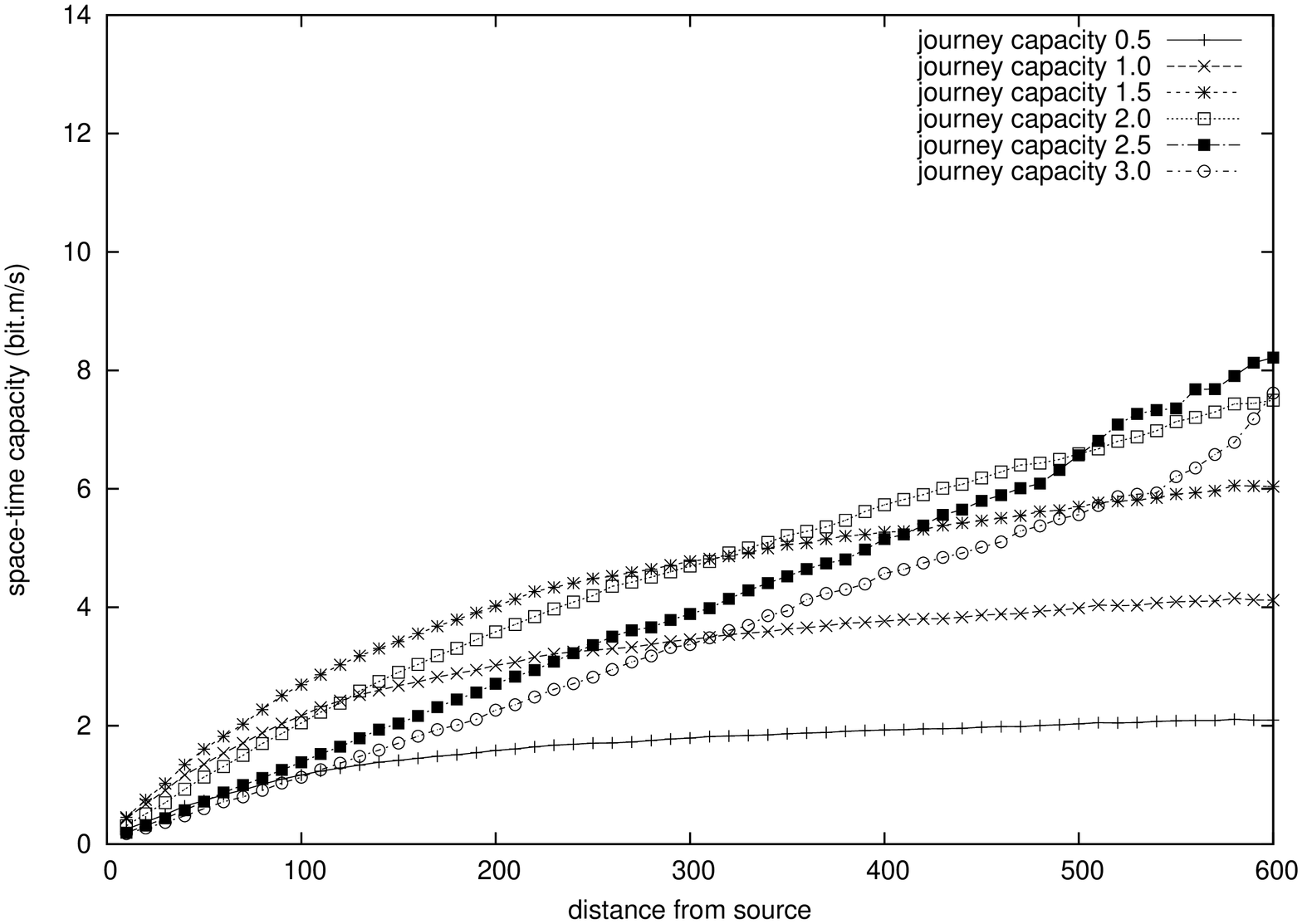}
\includegraphics[height=7.3cm,angle=270]{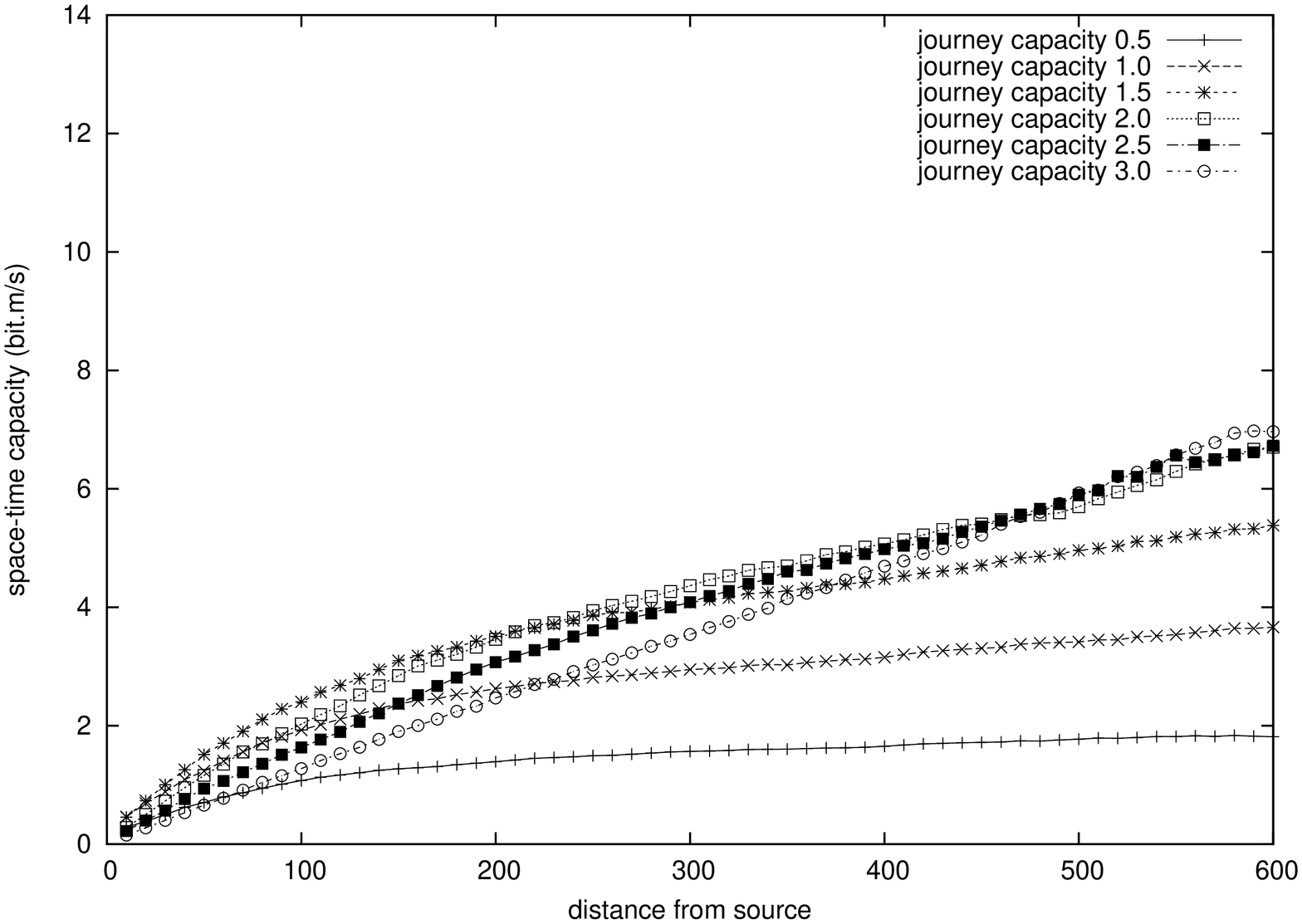}
\caption{Space-time capacity in bit-meters per second, versus distance from the source, for journey capacities $y=\{0.5;1;1.5;2;2.5;3\}$, with pure billiard mobility ($\tau=0$ $-$ top), and random walk mobility ($\tau=0.05$ $-$ bottom).}
\label{fig:spacetimecapacity}
\end{center}
\end{figure}

Finally, in Figure~\ref{fig:spacetimecapacity}, we plot the space-time capacity in bit-meters per second, versus the distance from the source, achieved by journeys of different capacities $y=\{0.5;1;1.5;2;2.5;3\}$, in the same simulation scenario.
The space-time capacity is obtained by multiplying the average propagation speed~$s(y)$ with the journey capacity~$y$.
We observe indeed that, for journey capacities up to $2$ units of data, the plots of the space-time capacity in Figure~\ref{fig:spacetimecapacity}, converge to $c(y)=s(y)y \approx vy$;
this is consistent with Remark~\ref{rem:bundle}.
For larger capacities, the space-time capacity has not converged to a constant value, due to the fact that the network domain is finite.
However, we note that, in a larger network, the space-time capacity would be larger for journeys of larger capacities.
In fact, in an infinite network, the space-time capacity would converge to a constant value for any finite journey capacity.

\section{Concluding Remarks}
\label{Sect:conclusion}

We characterized the space-time capacity limits of mobile DTNs, by providing lower (Theorem~\ref{theo:lower}) and upper bounds (Theorem~\ref{theo:upper}) on the information propagation speed, with a given journey capacity.
Moreover, we verified the accuracy of our bounds with extensive simulations
in several scenarios.

Such theoretical bounds are paramount in order to increase our understanding
of the fundamental properties and performance limits of DTNs, as well as to design or optimize the performance of specific routing protocols.
In fact, our results provide lower and upper bounds on the best achievable propagation delay of bundles of data, over large distances.

It is also worth noting that our analysis provides the first known lower bounds on the information propagation speed in mobile DTNs (for random waypoint-like mobility models), and generalize previously known upper bounds.

More specifically,
in the case of random waypoint-like mobility models, we showed that
for relatively large journey capacities, the information propagation speed is of the same order as the mobile node speed.
This implies that, in sparse but large-scale mobile DTNs, the space-time information propagation capacity in bit-meters per second remains proportional to the mobile node speed and to the size of the transported data bundles, when the bundles are relatively large.

\section*{Acknowledgment}

The authors would like to thank Matthieu Mangion for useful discussions that led to the proof of Lemma~\ref{lem:distT}.

\appendix

\subsection{Proof of Lemma~\ref{lem:meeting_rate} (Meeting Rate)}
When $n, \A \to \infty$,
we can consider an infinite network with a Poisson density of nodes $\nu=\frac{n}{\A}$ to simplify the proof.
In fact, if we consider an area $\A$ of the infinite network, the number of other nodes is given by a Poisson process of rate $n$, and we can \emph{depoissonize} it~\cite{js98},
to obtain the equivalent result when the number of nodes $n$ is large but not random.

Let $\bu$ be a unit vector always centered at the position of node $A$.
We denote by $f$ the rate at which mobile nodes enter the neighborhood range of node $A$ at position $R \bu$ with respect to the node location $\bz_A(t)$, where $R$ is the radio range.

Let us denote by $B$ a second network node, with a constant vector speed $\bv_B$.
The Poisson density of presence of $B$ at any location on the plane is $\nu$.
The relative speed of the nodes is $\bv_B-\bv_A$.
The projection of the relative speed on the vector $R \bu$ equals $\left(R \bu \cdot (\bv_B-\bv_A)\right)\bu$.
The rate at which any node $B$ enters the neighborhood range of the node $A$ at
$\bu$, is
$
f(\bv_A,\bv_B,\bu)= \max \{0, \bu \cdot (\bv_B-\bv_A)\nu R\}.
$

By averaging on~$\bu$, we have the total meeting rate:
$$
f(\bv_A,\bv_B) = \int_{-\frac{\pi}{2}}^{\frac{\pi}{2}} |\bv_B-\bv_A| \cos \psi \nu R d\psi = 2 \nu |\bv_B-\bv_A| R.
$$
Therefore, the rate at which a node meets new neighbors is proportional to their relative speed.
From the law of sines, the relative speed is proportional to $\sin(\frac{\Delta \psi}{2})$, where $\Delta \psi=\psi_1 - \psi_0$ is the angle formed between the speed vectors. By normalizing, we obtain the meeting rate.

\subsection{Proof of Lemma~\ref{lem:distT} (Distribution of Encounter Duration)}
We consider the encounter of two nodes $A$ and $B$, moving at speeds $\bv_A$ and $\bv_B$ respectively.
We define $\Delta \bv=\bv_B-\bv_A$ as the relative speed of the nodes.
Therefore, taking as a frame of reference the position of node $A$, node $B$ is moving at constant speed $\Delta \bv$, as illustrated in Figure~\ref{fig:chord}. We denote by $\Delta v$ the Euclidean norm of the relative speed (\ie the relative velocity).
From the law of cosines, it holds:
\begin{equation}
\Delta v = ||\bv_B-\bv_A||=2 v \sin(\frac{\psi}{2}),
\label{eq:relspeed}
\end{equation}
where $\psi\in [0,2\pi)$ is the angle
between the node speed vectors.

From Lemma~\ref{lem:meeting_rate}, the rate at which nodes meet is proportional to their relative speed. Therefore, normalizing~(\ref{eq:relspeed}), the angle $\psi$ is distributed according to the probability density function:
\begin{equation}
p_\psi(x) dx=\frac{1}{4} \sin(\frac{x}{2}) dx,~x\in [0,2\pi),
\label{eq:relangle}
\end{equation}
and, substituting $V=2v \sin(\frac{x}{2})$ according to~(\ref{eq:relspeed}), the density function $p_{\Delta v}(V)$ of the relative velocity is:
\begin{equation}
p_{\Delta v}(V) dV =\frac{V}{2 v} \frac{1}{\sqrt{4v^2-V^2}} dV,~V\in [0,2v].
\label{eq:relV}
\end{equation}

\begin{figure}[th!]
\centering
\includegraphics[width=4.5cm]{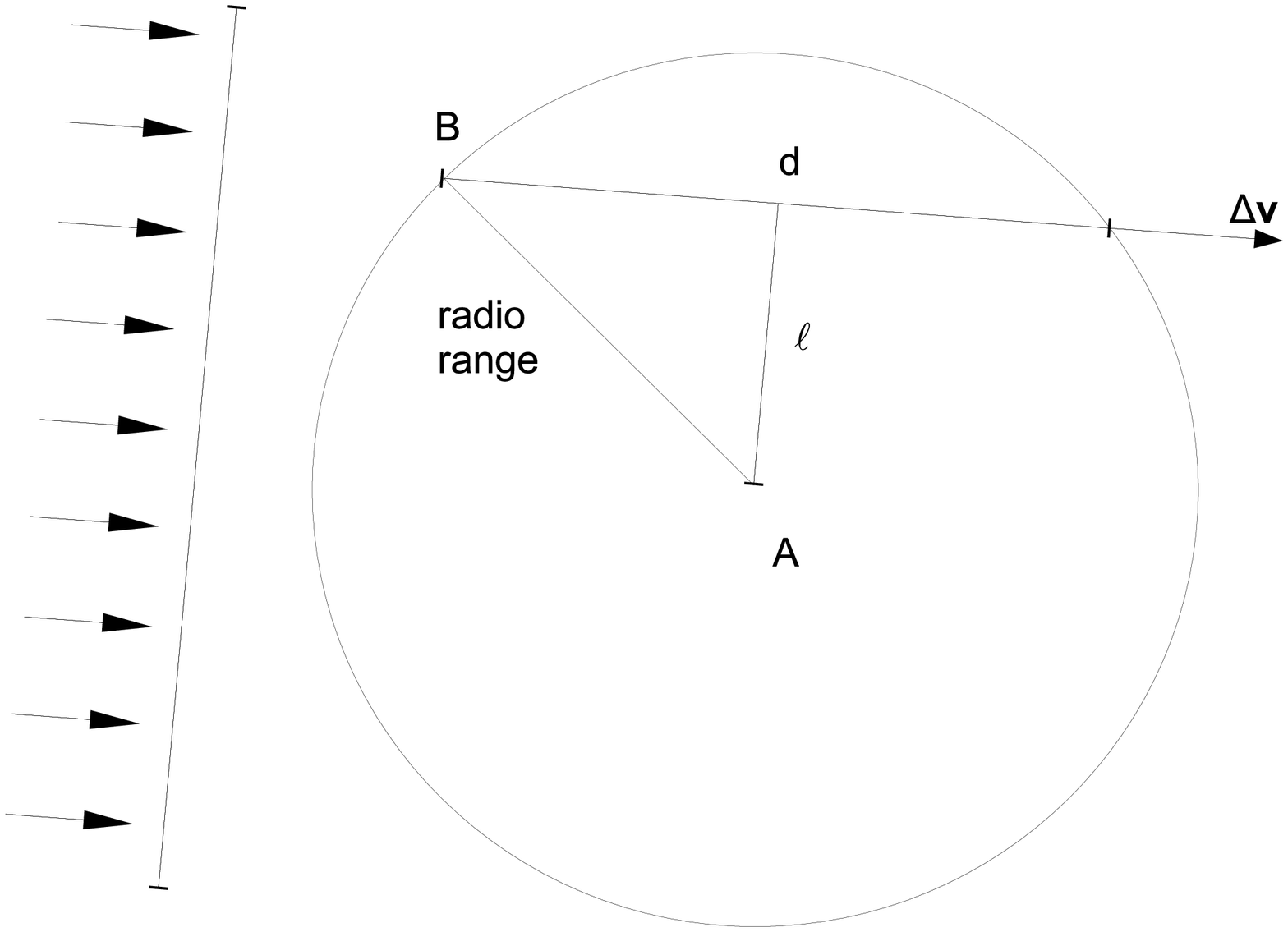}
\caption{Encounter of nodes $A$ and $B$ in the frame of reference centered at~$A$: $\Delta \bv$ is the relative speed of~$B$, $d$ is the length of the chord traveled by $B$ within range, $\ell$ is the distance of the chord $d$ from $A$.}
\label{fig:chord}
\end{figure}

\begin{figure}[th!]
\centering
\includegraphics[width=6cm]{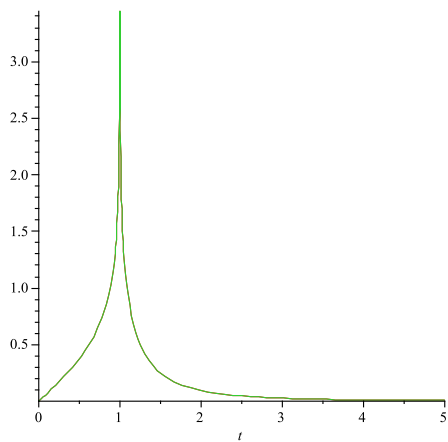}
\caption{Probability density function $p_T(t)=\frac{v}{4}\log\left| \frac{vt+1}{vt-1}\right| ( 1+\frac{1}{(vt)^2})-\frac{1}{2t}$ of the node encounter duration $T$, for $v=1$.}
\label{fig:pdfT}
\end{figure}

Always in the frame of reference of node~$A$, we denote by $d$ the distance traveled by node~$B$ within range of node~$A$. In other words, $d$ is the length of a chord of the circle of radius $R$ (the radio range), centered at node $A$.
We define $\ell$ as the distance of the chord from $A$, as depicted in Figure~\ref{fig:chord}.
We remark that, as a node moves and meets new neighbors, quantity $\ell$ is distributed uniformly at random between $0$ and $R$, since meetings occur equiprobably at any point of the diameter perpendicular to the node relative speed.
Therefore, since $d=2\sqrt{R^2-\ell^2}$, the distribution of the length $d$ is:
\begin{equation}
P(d>x) = \sqrt{1-\frac{x^2}{4R^2}}.
\label{eq:chord1}
\end{equation}
Differentiating, we obtain the probability density function:
\begin{equation}
p_d(x) =\frac{x}{2R \sqrt{4R^2-x^2}},~x\in[0,2R].
\label{eq:chord}
\end{equation}

If $T$ is the duration of the encounter, we have:
\begin{equation}
d=\Delta v \times T,
\label{eq:dvT}
\end{equation}
where all quantities are random variables.

Let us consider a given relative velocity $\Delta v=V$.
In this case, we can define the conditional probability density $p_T(t~|~\Delta v=V)$ of the encounter duration, with $t\in[0,\frac{2}{V}]$:
$$
p_T(t~|~\Delta v=V) dt = p_d(x~|~\Delta v=V) dx = p_d(Vt) V dt,
$$
where $x=Vt$, according to~(\ref{eq:dvT}).

Combining with~(\ref{eq:chord}),
\begin{equation}
p_T(t~|~\Delta v=V) = \frac{V^2 t}{2\sqrt{4-(Vt)^2}}.
\label{eq:condV}
\end{equation}

Considering the probability density function $p_T(t)$, and using~(\ref{eq:relV}) and~(\ref{eq:condV}), we have for $t~\geq 0$:
\begin{eqnarray*}
p_T(t)&=&\int_0^{2v} p_T(t~|~\Delta v=V) \times p_{\Delta v}(V)dV\\
	&=&\frac{1}{4}\log\left| \frac{\frac{v}{R}t+1}{\frac{v}{R}t-1}\right| \left( 1+\frac{R^2}{(vt)^2}\right)-\frac{R}{2vt}.
\end{eqnarray*}
We note that the fact that nodes bounce on the borders does not impact on this result.
We plot the probability density function $p_T(t)$ (for $R=1$, $v=1$) in Figure~\ref{fig:pdfT}.

By simple integration, we obtain the probability $P(T>t)$:
\begin{equation}
P(T>t)=\frac{1}{4}\log\left| \frac{\frac{v}{R}t+1}{\frac{v}{R}t-1}\right| \left( \frac{R}{vt}-\frac{v}{R}t\right)+\frac{1}{2}.
\label{cumT}
\end{equation}

For large $t$,
we have $\frac{\frac{v}{R}t+1}{\frac{v}{R}t-1}>0$. Therefore,
using the identity $\log x=2\sum_{n=0}^{\infty} \frac{1}{2n+1} \left( \frac{x-1}{x+1} \right) ^{2n+1}$, we have:
$$
P(T>t)
=\frac{R^2}{3(vt)^2}+ O(\frac{R^4}{(vt)^4}).
$$

\subsection{Proof of Lemma~\ref{lem:stage1} (Duration of Routing Stage 1)}

Since we consider meetings of relative angle at most $a$,
the relative speed of two meeting nodes is maximized when the angle between them is $a$ (and equals $2v\sin (\frac{a}{2})$).
Therefore, in order for the meeting duration $T$ to be at least equal to $y$, it is sufficient that the distance $d$ traveled within range, in the frame of reference of one of the nodes (see Figure~\ref{fig:chord}),
satisfies:
$$
d \geq vay \geq 2v\sin (\frac{a}{2}) y.
$$

\begin{figure}[t!]
\centering
\includegraphics[width=4.5cm]{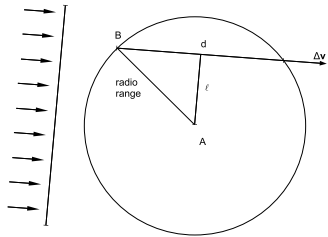}
\caption{Encounter of nodes $A$ and $B$ in the frame of reference centered at~$A$: $\Delta \bv$ is the relative speed of~$B$, $d$ is the length of the chord traveled by $B$ within range, $\ell$ is the distance of the chord $d$ from $A$.}
\label{fig:chord}
\end{figure}

According to~(\ref{eq:chord1}), $P(d>x)=\sqrt{1-\frac{x^2}{4}}$, and
$
P(T \geq y)
\geq \sqrt{1-\frac{a^2 v^2 y^2}{4}}.
$
Assuming that $y \geq \frac{1}{v \pi}$, we take $a=\frac{1}{2 v y}$,
$$
P(T \geq y) \geq \frac{\sqrt{15}}{4} \geq \frac{\pi}{4}.
$$
For smaller $y$, the same bound clearly still holds.

From Lemma~\ref{lem:meeting_rate}, the probability to meet a node at an angle in $[\frac{a}{2},a]$ is $P_{a}=\cos(\frac{a}{4})-\cos(\frac{a}{2}) \geq \frac{1}{16}a^2,$
since $a \leq \frac{\pi}{2}$.

The
rate at which a node meets new nodes at such an angle, ensuring that the meeting duration is at least $y$, is:
$$
f_1 \geq \frac{4v \nu}{\pi} P_{a} P(T \geq y)
\geq \frac{v \nu}{16} a^2.
$$

We note that the angle $\phi_C$ determines the distance $d_{B}$ of node $C$ trajectory from the point $B$ (see Figure~\ref{meetingpoint}).
In fact, it holds:
$
d_{B}=|CB| \sin \phi_C.
$
When a node moves, $\phi_C$ varies, while $d_{B}$ remains unchanged.
In fact $\phi_C$ always increases when a node moves towards the destination.
However, after a node movement of distance $\delta$, we have $\Delta \phi_C=O(\frac{\delta}{|CB|})$, and if $\delta=o(|CB|)$,
$\phi_C$ is not modified asymptotically.

Thus, if the initial angle between the source and the destination is $b$,
the expected time $E(t_1')$ until $\frac{a}{2} \leq \phi_C \leq a$ is:
$$
E(t_1') \leq \frac{2 b}{af}
\leq \frac{32 \pi}{a^3 v \nu}
=\Theta(\frac{1}{v \nu}).$$

From Lemma~\ref{lem:meeting_rate}, the rate at which a node meets nodes at relative angle $[\psi,\psi+d\psi]$ is $\frac{2 v \nu}{\pi} \sin (\frac{\beta}{2}) d\beta$.
Therefore, the node $C$ that last received the information meets new nodes $C'$ with angle $\phi_C' \leq \frac{1}{\sqrt{r}}$, and with meeting duration at least $y$, with
rate (assuming that $\phi_C$ remains between $\frac{a}{2}$ and $a$):
{\small
$$
f_2 \geq \frac{2 v \nu}{\pi} P(T>y) \int_{a-\frac{1}{\sqrt{r}}}^{a+\frac{1}{\sqrt{r}}} \sin (\frac{a}{4}+x) dx
\geq \frac{v \nu \sin (\frac{a}{4})}{4 \sqrt{r}} +O(r^{-\frac{3}{2}}).
$$}
and the expected time $E(t_1'')$ until meeting such a node is $\Theta(\frac{\sqrt{r}}{v \nu})$ (we note that $\frac{1}{a} \leq 2 K$).
We notice that the $t_1''=o(r)$ almost surely, and we can indeed assume that $\phi_C$ remains constant until meeting $C'$.

Therefore,
it holds that the duration $t_1$ of stage 1 is
$t_1=t_1'+t_1''=O(\frac{\sqrt{r}}{v \nu})$ almost surely.
The distance traveled is $v t_1 + O(\frac{1}{a})$, where the second term corresponds to the further distance moved by the information in $O(\frac{1}{a})$ transmissions.
Since $\frac{1}{a}=O(1)$,
the total distance traveled is
$O(\frac{\sqrt{r}}{\nu})$.

\subsection{Proof of Lemma~\ref{lem:stage3} (Duration of Routing Stage 3)}

We proceed equivalently to stage~1.
Stage~3 ends when a node with angle $\phi_C \leq \frac{1}{2 r_C}$ receives the information.
Equivalently to the proof of Lemma~\ref{lem:stage1}, the expected time $t_3'$ until the relative speed of the node to the rendez-vous point $A$ is between $\frac{a}{2}$ and $a$ is
$E(t_3')=\Theta(\frac{1}{v \nu})$.

We consider meetings with nodes $C'$, such that $2 r_{C'} \leq  \frac{\sqrt{r}}{k_1}$, where $k_1>0$ is a constant.
The node $C$ that last received the information meets new nodes $C'$ with angle $\phi_{C'} \leq \frac{k_1}{\sqrt{r}}$ ($\leq \frac{1}{2 r_{C'}}$), and with meeting duration at least $y$, with
rate (assuming that $\phi_C$ is between $\frac{a}{2}$ and $a$):
{\small
$$
f_2 \geq \frac{2 v \nu}{\pi} P(T>y) \int_{a-\frac{k_1}{\sqrt{r}}}^{a+\frac{k_1}{\sqrt{r}}} \sin (\frac{a}{4}+x) dx
\geq \frac{k_1 v \nu \sin (\frac{a}{4})}{4 \sqrt{r}} +O(r^{-\frac{3}{2}}).
$$}
and the expected time $E(t_3'')$ until meeting such a node is $\Theta(\frac{\sqrt{r}}{v \nu})$.
Since $\phi_C$ varies, if it becomes larger than $a$ (or smaller than $\frac{a}{2}$), the information is forwarded to a new node such that $\phi_C$ is between $\frac{a}{2}$ and $a$ again (in constant time).

Moreover, we have indeed that $r_C=O(\sqrt{r}) \leq \frac{\sqrt{r}}{2 k_1}(1+O(1))$ for some positive constant $k_1$, since the distance traveled at this stage is at most $vt_3'+vt_3''+O(\frac{1}{a})=O(\sqrt{r})$.
We assume that $r_C \leq r_B$, which we can ensure by choosing point $B$ sufficiently far from the rendez-vous point $A$.
In this case, when the information is transmitted to node $C$, the node's direction, with respect to the destination's speed, is of angle at most $\phi_C \leq a$.
Therefore, after time $t_3=\Theta(\frac{\sqrt{r}}{v \nu})$, the destination is reached with probability strictly larger than $0$.

\subsection{Proof of Corollary~\ref{cor:randomwalk}}
W. l. o. g., we take $R=1$ and $G=1$.
Let $(\rho,\theta(\rho))$ be an element of the set $\CK$. 
We have:
\begin{equation}
\theta(\rho)=\sqrt{(\tau+\gamma(y) \nu H(\rho))^2+\rho^2 v^2}-\tau,
\label{thetarho}
\end{equation}
with 
\begin{equation}
H(\rho)=\frac{4 \pi v I_0(\rho)}{1-\gamma(y) \frac{\pi \nu}{2 \rho} I_1(\rho)}.
\label{Hrho}
\end{equation}
For $y$ sufficiently large,  such that $\gamma(y) \nu = \frac{\pi^2 \nu}{8 v y} \to 0$, 
$$\theta(\rho)=\sqrt{\tau^2+\rho^2 v^2}-\tau+\frac{\tau}{\sqrt{\tau^2+\rho^2 v^2}}H(\rho)\frac{\pi^2 \nu}{8 v y} +O(\frac{\nu^2}{y^2}),$$
and, since $H(\rho)=4 \pi v I_0(\rho)+O(\frac{\nu^2}{y^2})$, we obtain the ratio:
$$
\frac{\theta(\rho)}{\rho}=\frac{\sqrt{\tau^2+\rho^2 v^2}-\tau}{\rho}+\frac{\tau}{\sqrt{\tau^2+\rho^2 v^2}} I_0(\rho)\frac{\pi^3 \nu}{2 y \rho} +O(\frac{\nu^2}{y^2 \rho}).
$$

Therefore, when $\rho\to 0$,
$$
\frac{\theta(\rho)}{\rho}=\frac{\rho v^2}{2\tau}+\frac{\pi^3 \nu}{2 y \rho}+O(\frac{\nu^2}{y^2 \rho}+\frac{\nu \rho^2}{y})
$$

The sum $\frac{\rho v^2}{2\tau}+\frac{\pi^3 \nu}{2 y \rho}$ is minimized when $\rho=\frac{\pi}{v} \sqrt{\frac{\nu \tau}{y}}$, and its minimum is $\pi v \sqrt{\frac{\nu}{y \tau}}$.

As a result, the ratio $\frac{\theta(\rho)}{\rho}$ is minimized with value $\pi v \sqrt{\frac{\nu}{y \tau}} + O(\left(\frac{\nu}{y}\right)^{\frac{3}{2}})$, which corresponds to the propagation speed bound.

\subsection{Proof of Corollary~\ref{cor:billiard}}

Again, we take w. l. o. g., $R=1$ and $G=1$, and we consider the kernel set $(\rho,\theta(\rho))$.
From~(\ref{thetarho}), when $\tau \to 0$,
$$
\theta(\rho)=\sqrt{(\gamma(y) \nu H(\rho))^2+\rho^2 v^2}+O(\tau).
$$

We obtain the ratio:
$$
\frac{\theta(\rho)}{\rho}=\sqrt{\frac{(\gamma(y) \nu H(\rho))^2}{\rho^2}+ v^2}+O(\frac{\tau}{\rho}).
$$

In this case, $\sqrt{\frac{(\gamma(y) \nu H(\rho))^2}{\rho^2}+ v^2}$ is minimized when the quantity $J(\rho)=\frac{\gamma(y) \nu H(\rho)}{\rho}$ is also minimized. 

We take $\gamma(y) = \frac{\pi^2}{8 v y}$, since this is an upper bound for any value of the parameters.
Thus, using~(\ref{Hrho}) when $\frac{\nu}{y} \to 0$,
$$
J(\rho)=\frac{\pi^2 \nu H(\rho)}{8 v y \rho}=\frac{\pi^2 \nu I_0(\rho)}{8 y \rho}+O(\frac{\nu^2}{y^2}).
$$
Therefore, the minimum of $J(\rho)$ is $\frac{\pi^2 \nu}{8 y} \min_\rho (\frac{I_0(\rho)}{\rho})+O(\frac{\nu^2}{y^2})$, attained for $\rho=1.608\ldots$, and we have the propagation speed upper bound: $v+O(\frac{\nu^2}{y^2}+\tau)$.

\end{document}